%% file: bmwpaper.tex
\documentclass[article]{aa}
\usepackage{psfig,latexsym,longtable,lscape,supertabular}
\usepackage[authoryear]{natbib}

\def\x{X-ray }
\def\lx{L$_X$ }
\def\Lx{L$_X$}
\def\lb{L$_B$ }
\def\Lb{L$_B$}
\def\ergsec{erg s$^{-1}$}
\def\ergcmsec{erg cm$^{-2}$ s$^{-1}$}

\begin{document}

\pagenumbering{arabic}
\title{A sample of X-ray emitting normal galaxies from the BMW -- HRI Catalogue}

\author{Marzia Tajer\inst{1}$^,$ \inst{2}, Ginevra Trinchieri\inst{1}, 
Anna Wolter\inst{1}, Sergio Campana\inst{1}, Alberto Moretti\inst{1}, 
Gianpiero Tagliaferri\inst{1}}
  
\institute{
INAF-Osservatorio Astronomico di Brera, via Brera 28, 20121
Milano Italy
\and
Universit\`a di Milano - Bicocca, Dipartimento di Fisica, P.za della Scienza 3, 
20126 Milano Italy
}
\offprints{M.~Tajer}
\mail{tajer@brera.mi.astro.it}

\date{Draft: \today}
\authorrunning{Tajer et al.}
\titlerunning{Galaxies in BMW -- HRI Catalogue}

\abstract{We have obtained a sample of 143 normal galaxies 
with \x luminosity in the range $10^{38} - 10^{43}$ \ergsec{} 
from the cross-correlation of the 
ROSAT HRI Brera Multi-scale Wavelet (BMW -- HRI) Catalogue
with the Lyon-Meudon Extragalactic Database (LEDA). We find
that the average \x properties of this sample are in good
agreement with those of other samples of galaxies in the
literature. We have selected a complete flux limited
serendipitous sample of 32 galaxies from which we have
derived the logN-logS distribution of normal galaxies in the
flux range $1.1 - 110 \times 10^{-14}$ \ergcmsec. The resulting
distribution is consistent with the euclidean $-1.5$ slope.
Comparisons with other samples, such as the Extended Medium
Sensitivity Survey, the ROSAT All Sky Survey, the XMM - Newton/2dF
survey and the \textit{Chandra} Deep Field Survey indicate that
the logN-logS distribution of normal galaxies is consistent
with an euclidean slope over a flux range of about 6 decades. 
 }

\maketitle

\section{Introduction}
Detailed \x studies of normal galaxies have been possible only with the
advent of imaging instruments aboard the \textit{Einstein} observatory.
The \textit{Einstein} results were summarized in the catalogue and
atlas published by \citet{Fabbiano etal 92}, who constructed a large
homogeneous sample of 493 galaxies, included in either A
Revised Shapley-Ames Catalog of Bright Galaxies \citep{Sandage&Tamman 81}, 
or in the Second Revised Catalog of Bright Galaxies
\citep{deVaucouleurs etal 76}, both targets and serendipitously
detected in \textit{Einstein} fields. While representative of the
galaxy population \citep[see e.g.][]{Shapley etal 01, Eskridge etal 95}, 
it was not constructed to be a complete, unbiased sample, and it
is likely to contain unknown selection biases.

Elliptical galaxies were found to retain large amounts ($10^8 -
10^{11}$ M$_{\sun}$) of hot gas (T $\sim 10^7$ K) whose thermal
emission dominates their \x luminosities, while in normal spirals the
integrated contribution of the evolved stellar sources, such as
supernova remnants and \x binaries, is generally the dominant component
\citep[see][]{Fabbiano 89, Fabbiano etal 92, Kim etal 92}. Extended
emission from a hot, gaseous component in spiral galaxies was detected
only in some cases \citep{Fabbiano&Trinchieri 87, Vogler&Pietsch 96,
Trinchieri etal 88}, or associated with starburst activity.

Subsequent observations of individual sources made by \textit{ROSAT}
and \textit{ASCA} confirmed most of the \textit{Einstein} results and
added interesting information on the \x properties of normal galaxies
\citep[see among others][]{Roberts&Warwick 00, Read etal 97,
Brown&Bregman 98, Beuing etal 99} in the local universe.

With the launch of \textit{XMM - Newton} and \textit{Chandra},
the study of the \x properties of ``normal'' galaxies at
intermediate ($z \sim 0.1$) or cosmological distances 
\citep{Brandt etal 01, Hornschemeier etal 02, Hornschemeier etal 03, 
Georgakakis etal 03, Georgakakis etal 04a, Georgakakis etal 04b, Norman etal 04} 
was made possible, thanks to significantly
improved sensitivity, spatial and spectral resolution of the
instruments. In spite of the large number of papers, however, a truly complete
sample of \x emitting normal galaxies in the local universe with a significant 
number of objects has not been properly discussed  in the literature so far. 
\citet{Georgakakis etal 03} give galaxy number density at 
$F_{0.5-2} \sim 7 \times 10^{-16}$ \ergcmsec{} from stacking analysis, 
while \citet{Georgakakis etal 04a} present a sample of 26 \x sources 
detected in an area of $\sim 2.5$ deg$^{-2}$, of which only 2 
are however classified as normal galaxies. Only recently 
\citet{Georgakakis etal 04b} have presented a larger sample 
of 11 normal galaxies detected in an area of $\sim 4.5$ deg$^{-2}$.

Two other samples of ``normal'' galaxies are available in the
literature, selected in the \textit{Chandra} Deep Fields
\citep{Hornschemeier etal 03, Norman etal 04}. However their median
redshifts ($z = 0.297$, \citet{Hornschemeier etal 03} and $z = 0$ to
$z = 1.3$, \citet{Norman etal 04}) indicates that they should not be
considered as ``local". 

The large database provided by ROSAT has been exploited only marginally
to derive unbiased and complete sample of galaxies. \citet{Zimmermann etal 01}
 have selected a sample of \textit{candidate normal galaxies}
from  the \textit{ROSAT} All Sky Survey (RASS) Bright Source Catalogue
\citep{Voges etal 99} above a flux limit of about $10^{-12}$ \ergcmsec{} 
($0.1 - 2.4$ keV band).
A few samples have been derived from the \textit{ROSAT} Position
Sensitive Proportional Counter (PSPC) pointed observations such as the
WGA \citep{White etal 94} and the ROSPSPC (\textit{ROSAT} team 2001),
and a new catalogue of galaxies is in progress (G. Peres, private
communication). Here we exploit the potential provided by the
Brera Multi-scale Wavelet (BMW -- HRI) catalogue \citep{Panzera etal 03}
to extract a sample of normal galaxies, as we discuss in the next
sections. While the PSPC is probably more efficient at detecting faint
and extended sources such as galaxies, the sharp core of the HRI point
spread function allows the detection of sources in more crowded fields
and to establish the extension for bright small-size sources, providing
a good complement to the PSPC data.
  
\section{The sample} \label{sample}
In order to create a complete, serendipitous sample of galaxies with \x
emission, we made use of \x data from the BMW \textit{ROSAT} HRI
catalogue and optical data from LEDA (Lyon-Meudon Extragalactic Database). 
The BMW -- HRI catalogue consists
of 29089 \x sources detected in 4303 \textit{ROSAT} HRI pointed fields
with exposure times longer than 100 s using a multiscale wavelet
algorithm \citep{Lazzati etal 99, Campana etal 99, Panzera etal 03}.
Sources detected with a significance $\geq$ 4.2 $\sigma$ are contained in the
catalogue, that provides name, position, count rate, flux and
extension along with the relative errors. In our study we used the full
catalogue, but we excluded \x sources in the Trapezium field, which is
a rich stellar cluster in the Milky Way, where the high density of the X-ray
sources would prevent proper optical identifications.
The BMW -- HRI catalogue
can be searched 
via the HEASARC Browse\footnote{\tt http://heasarc.gsfc.nasa.gov/cgi-bin/W3Browse/
w3browse.pl}
or via the Brera Observatory web site.\footnote{\tt http://www.merate.mi.astro.it:8081/interroga/
dbServer?cmd=bmw2} 
 
Created in 1983 at Lyon Observatory, LEDA\footnote{There exist several
mirrors to access LEDA; we used the OAB one,
\tt{http://www.brera.mi.astro.it/hypercat/}} has been the first
database of extragalactic objects and it is continuously updated. It
gives a free access to the main astrophysical parameters (coordinates,
morphological type, diameter and axis ratio, apparent magnitudes and
colors, radial velocity, surface brightness, etc) for about $10^6$
galaxies over the whole sky. The completeness in apparent B - magnitude
is satisfied up to m$_B$ = 15.5 \citep[see][]{Paturel etal 97}.

To obtain a representative sample of galaxies we started from the BMW
-- HRI catalogue and included only serendipitous detections, avoiding
the targets. We chose a 3\arcmin{} radius to define the typical region
of the target, and selected only sources at off-axis angles $\theta >
3$\arcmin. In spite of their off-axis location, we had to exclude 12
additional sources that were targets of the observations.  We then
cross - correlated the positions of the \x sources in the BMW -- HRI
catalogue with those of galaxies present in the LEDA database version
of 1999, with a tolerance of 20\arcsec, which should be a reasonable
guess to detect extended objects like galaxies and to avoid most of
chance coincidences.  This criterium is not appropriate for very
extended galaxies (like e.g. M~31 or M~33), where a large number of
sources is detected at distances significantly larger than our
tolerance radius. Therefore we could be selecting against
large size galaxies, if there is no source within a distance of 
$\sim 20''$ from the nucleus.  However, this should not be a
concern in this study, since the surface density of large
galaxies is small: in particular, in the LEDA catalog, the
density of galaxies with D$_{25} >3'$ is  $< 3 \times 10^{-2}$
deg$^{-2}$, which implies about $< 10$ in the area we surveyed. 
 
\begin{figure}
\psfig{figure=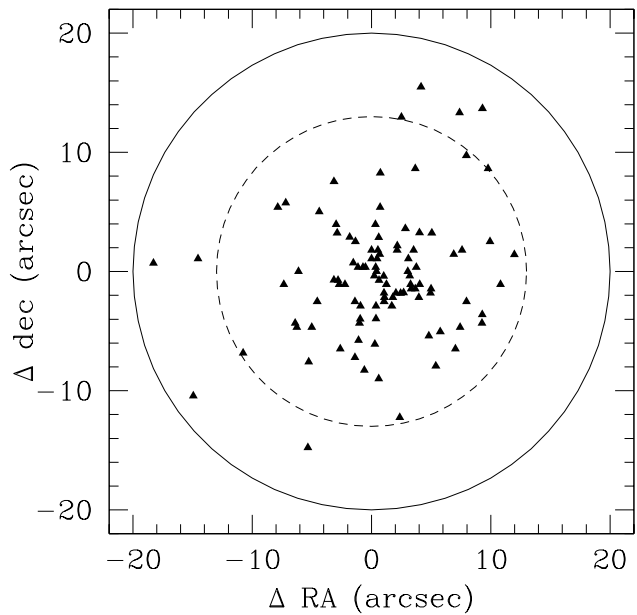,width=8.8cm,clip=}
\caption{Displacement between optical and \x positions for point-like sources in
the total sample. The solid circle refers to the cross - correlation radius of
20\arcsec; the dashed one, with 13\arcsec{} radius, contains about 90\% of the
identifications.}
\label{deltaxott}
\end{figure}

To check the goodness of our choice, we plot in Fig. \ref{deltaxott}
the relative shifts between optical and \x positions. 
We use only point-like \x sources because in extended sources the association
with a single optical object could be misleading (e.g. in groups and pairs).
About 90\% of identifications are within 13\arcsec{} (represented by the
dashed circle in Fig. \ref{deltaxott}).

This is in good agreement with the HRI positional 
uncertainty, since the best attitude solution guarantees that on average 
known objects are detected within 10\arcsec{} of their catalog position,
although with possible additional discrepancy mainly in declination 
(see the ROSAT Handbook e.g. at {\tt
http://heasarc.gsfc.nasa.gov/docs/rosat/ruh/
handbook/handbook.html}).
We verified that all \x sources with X-ray/optical off-set $>$ 
10\arcsec{} are still within the galaxy (i.e. D$_{25}$). 
We conclude that the association of the X-ray sources with galaxies found 
by the cross-correlation is sound on positional grounds.

The cross-correlation yields 399 \x sources associated with 281
galaxies. Being constructed field-by-field, the BMW -- HRI catalogue
contains multiple detections of the same source. Therefore we only have
283 distinct \x sources associated with galaxies (two distinct sources but of 
different extent are
associated with both NGC~1399 and M~86; we list them in Table \ref{sampletab}, 
but we consider only the largest one for computing fluxes and luminosities).
 
To check the results of the cross-correlation, we inspected HRI images
for each source and obtained information from the NASA/IPAC
Extragalactic Database (NED) to a) eliminate spurious coincidences [15
sources], b) eliminate AGNs [47 objects] and c) select clusters [57
objects].

\begin{enumerate}
\renewcommand{\labelenumi}{\alph{enumi}}
\item. We eliminated a source when the obvious optical counterpart was
not a LEDA galaxy, but a background or foreground object. This was
checked both using NED and the X-ray/optical contour maps.

\item. We excluded objects classified in \citet{Veron 01 cat} as
Seyfert~1 galaxies, QSOs, BL Lacs or AGNs. For Seyfert 2 galaxies, the
\x emission could have a non nuclear origin, so we eliminated them when
we saw from images that the \x emission was point-like and well
centered on the nucleus, while we retained those for which we found an
extended emission.\\ 
We also retained starburst galaxies, LINERs and
objects that we know from the literature could have a nuclear source
but also non - nuclear \x emission (e.g. NGC~3079).

\item. Several galaxies lie inside a cluster: we inspected \x images and
we eliminated those objects whose \x emission was indistinguishable
from the cluster's, but we retained galaxies for which emission clearly
associated with the galaxy is detected above the cluster background 
(see e.g. PGC 12350 in Fig.
\ref{atlasfig}).\\  
We retained galaxies in poor groups even when
they are the brightest member since there is still ambiguity in the
literature between emission from bright early-type galaxies and
groups, which are often analyzed in the same context
\citep[see e.g.][]{Mamon 92, Dell'Antonio etal 94, Pildis etal 95,
Ponman etal 96, Mahdavi etal 97, Mulchaey&Zabludoff 98, Helsdon&Ponman 00, 
Mulchaey etal 03, Osmond&Ponman 04, Helsdon&Ponman 03, Jones etal 03}. 
\end{enumerate}

Moreover, in 5 cases the same \x source was associated with two or more
galaxies in a pair or in a group: since we could not discriminate on a
positional basis, we chose the brightest galaxy in the pair or in the
group. 

These selection criteria yielded a total of 195 \x sources (including 
multiple detections) associated with 143 galaxies 
whose properties are listed in Table \ref{sampletab}, available in 
electronic form. We report here the first page as an example.
Column (1) gives the BMW -- HRI name of the source, columns (2) and (3)
the position of the \x peak, column (4) labels the source as point-like 
(p) or extended (e), column (5) gives the radius in which we
computed count rates for extended sources (see section~\ref{data}), column (6) 
the major semiaxis of the ellipse, for extended \x sources
for which count rates have been computed in elliptical regions 
(see section~\ref{data}), column (7) the minor semiaxis of the ellipse,
column (8) the LEDA galaxy associated to the \x source,
column (9) the name of the galaxy in other common catalogues (e.g. NGC),
column (10) the morphological type, column (11) the distance of galaxy,
column (12) the apparent B - magnitude corrected for galactic
extinction, inclination and redshift effects \citep[see][]{Paturel etal 97}, 
columns (13) and (14) the HRI count rate and error, columns
(15) and (16) the \x flux and error ($0.1 - 2$ keV; count rates and
fluxes have been recomputed with respect to values reported in the BMW
-- HRI catalogue; see section \ref{data} for details), column (17) 
the logarithm of \x luminosity  and column
(18) specifies whether the galaxy is in the complete subsample (c), 
in cluster (Cl) or in group (Gr).

Information about magnitudes and redshifts are from LEDA and NED except
for 3 objects that we observed ourselves (see section \ref{optical}).
We calculated distances from redshifts, assuming $H_0$ = 50 km s$^{-1}$
Mpc$^{-1}$; when the heliocentric radial velocity of galaxy was less
than 3000 km s$^{-1}$, however, we used distances from Nearby Galaxy
Catalogue \citep{Tully 88} corrected for $H_0$ = 50 km s$^{-1}$
Mpc$^{-1}$.

One hundred and sixteen of the 143 galaxies have redshift and magnitude
(including our observations). Nineteen galaxies have m$_B$ and no $z$,
3 have $z$ and no m$_B$ and 5 have neither $z$ nor m$_B$. The reshift 
distribution ranges from $z = 0$ from $z \sim 0.15$, but $\sim$ 90\% of 
galaxies have $z < 0.07$.

\clearpage
\begin{landscape}
\begin{table}
\begin{center}
  \fontsize{8pt}{10pt}
  \begin{verbatim}

(1)                (2)         (3)        (4) (5) (6) (7)  (8)         (9)                           (10)  (11) (12)    (13)  (14)   (15)  (16) (17)  (18)
----------------------------------------------------------------------------------------------------------------------------------------------------------------   
BMW000523.9+161307 00 05 23.83 +16 13 11.3 e  52           PGC 0000372                                     698          15.74  2.49   7.40  1.17 43.59 Cl    
BMW002055.1+215208 00 20 55.43 +21 51 52.2 p               PGC 0001333 IC 1543                       Sbc   112  14.54    7.63  1.11   3.58  0.52 41.71 c     
BMW002055.2+215208                        		   	        		      	     	              	        
BMW002549.3-453227 00 25 49.26 -45 32 26.5 p               PGC 0143535                               Sab        16.57   18.64  3.16   7.64  1.30             
BMW002950.2-405630 00 29 50.99 -40 56 37.9 p               PGC 0130966 DUKST 294-9                   Sc    241  15.35   19.06  2.31   7.81  0.95 42.72 c     
BMW003652.3-333310 00 36 52.82 -33 33 14.7 p               PGC 0002204 ESO 350-IG38                  S?    123  14.96    2.74  0.43   1.12  0.14 41.27 c
BMW003918.5+030220 00 39 18.47 +03 02 14.8 p               PGC 0002362 NGC 194                       E     103  13.09   14.75  1.94   6.49  0.85 41.89 c; Gr
BMW003948.3+032219 00 39 48.77 +03 22 21.0 p               PGC 0002401 UM 57                                             6.15  0.49   2.70  0.22         
BMW004242.0+405154 00 42 42.48 +40 51 52.6 e  60           PGC 0002555 NGC 221                       E       1   8.18   52.44  1.06  26.75  0.54 38.52 Gr
BMW010716.2+323117 01 07 16.20 +32 31 16.6 p               PGC 0003966 NGC 379                       S0    110  13.48    1.37  0.35   0.67  0.17 40.98 Gr
BMW010717.4+322857 01 07 17.76 +32 28 57.0 p               PGC 0003969 NGC 380                       E      88  13.24    4.19  0.50   2.06  0.24 41.26 Gr
BMW010717.8+322857 	                  		   	        		      	     	              	        
\end{verbatim}
\end{center}
\normalsize
\begin{list}{}{}
\item[(1)] BMW -- HRI source name
\item[(2)] \x coordinates: RA (J2000)
\item[(3)] \x coordinates: Dec (J2000)
\item[(4)] \x source point-like (p) or extended (e)
\item[(5)] Radius of circle in which count rates were computed, for extended \x sources
\item[(6)] Major semiaxis of ellipse in which count rates were computed, for some peculiar extended \x sources
\item[(7)] Minor semiaxis of ellipse in which count rates were computed, for some peculiar extended \x sources
\item[(8)] LEDA galaxy associated
\item[(9)] Other galaxy name
\item[(10)] Morphological galaxy type
\item[(11)] Galaxy distance (Mpc)
\item[(12)] Apparent B - magnitude
\item[(13)] \x count rate (10$^{-3}$ count s$^{-1})$
\item[(14)] \x count rate error (10$^{-3}$ count s$^{-1})$
\item[(15)] \x flux ($0.1 - 2$ keV band; $10^{-13}$ \ergcmsec)
\item[(16)] \x flux error ($0.1 - 2$ keV band; $10^{-13}$ \ergcmsec)
\item[(17)] Logarithm of \x luminosity ($0.1 - 2$ keV band; \ergsec)
\item[(18)] Notes: c if the galaxy is in the complete subsample, Cl if in cluster, Gr if in group
\end{list}
\caption{First page of Table of the total sample, available in electronic form.}
\label{sampletab}
\end{table}
\end{landscape}

\subsection{The complete serendipitous subsample} \label{complete} 
To study the general properties of the sample we need to derive a subsample
with well known completeness criteria and limits. To this end we
constructed a complete sample with both \x and optical flux limits. We must 
consider both
the \x and the optical completeness criteria. The \x completeness is
related to the BMW -- HRI catalogue, that includes all sources with a
significance $\geq$ 4.2 $\sigma$. To take into account the optical
limits, we excluded galaxies fainter than m$_B$ = 15.5
\citep[the completeness limit assumed for the LEDA Catalogue, see][]{Paturel etal 97}.

We have also excluded objects at low galactic latitude ($|b| \leq$
10\degr) to avoid source confusion in the galactic plane and \x sources
with off-axis angle $\theta \ge 18$\arcmin{} to 
match the circular HRI field of view used in the sky coverage computation
(see section \ref{lognlogs}). 

The resulting sample of 96 objects is complete both in \x and
optical, at the given limits. However, since galaxies are often in 
agglomerates, some of them
are expected to be related to the targets
and therefore not truly
serendipitous. We therefore excluded all sources known to be associated
with the target (e.g. galaxies in pairs, groups or clusters; 52 objects).
When an association was not documented (e.g. from
NED, LEDA) we conservatively excluded galaxies at the same redshift of
the target (12 objects).

The complete, serendipitous sample of 32 galaxies thus obtained is given 
in Table \ref{compltab} and will be
used to calculate the logN-logS distribution in the local universe ($z < 0.07$), 
as will be described in detail in
section \ref{lognlogs}.

\section{\x characterization} \label{data}
The BMW -- HRI catalogue provides count rates derived  using the wavelet
algorithm, in an automated way and under particular
assumptions \citep[for details, see][]{Lazzati etal 99, Campana etal
99}. We verified that,
while count rates are correctly computed for point-like sources, for
very extended sources such as NGC~1399, the extension and count rates given by the
algorithm are underestimated. Moreover, when multiple observations are
available we can improve the statistics by considering the full set of
data. We have therefore recalculated all count rates, using the BMW -- HRI
positions and original HRI data retrieved from the public archives\footnote{{\tt
http://wave.xray.mpe.mpg.de/} and {\tt http://heasarc.gsfc.nasa.gov/}} , in
pulse height analyzer (PHA) channels $1 - 10$ to increase the
signal-to-noise ratio 
\citep[for a justification of this choice of PHA, see][]{Trinchieri etal 97}. 
When multiple observations of the same
field are available, we summed the data if they have the same pointing
coordinates and comparable exposure times. Otherwise, we typically used
the longer exposures or those where the source is closer to the field's
center.

\subsection{Count rates} 
We classified sources as point-like or
extended, based on the radial distribution of the emission relative to
the shape of the HRI point spread function 
\citep[PSF; for a description of the \textit{ROSAT} PSF see][]{Boese 00} at the
corresponding off-axis angle. For the ``extended sources'' the counts 
are taken from the largest
region that contains source counts, generally a circle of radius 
reported in Table \ref{sampletab}, evaluated from the radial profile. 
In few cases, given the particular shape of the source, we used 
ellipses to evaluate the counts; we give in Table \ref{sampletab} minor 
and major semiaxis. There are also a few extended sources for which 
ad hoc regions (neither circles nor
ellipses) have been used to estimate counts: these are PGC~5323, PGC~5324,
PGC~13418, PGC~13433 and PGC~56962.   

For point-like sources the counts are obtained in a circular region
centered at the peak of \x emission, with radius that includes about
90\% of source counts according to the PSF. The PSF degrades as the
angular distance from the center of field increases, so we chose a
radius of r = 18\arcsec{} for 3\arcmin $ < \theta \leq 10$\arcmin, r =
25\arcsec{} for 10\arcmin $< \theta \leq 15$\arcmin{} and r =
40\arcsec{} for $\theta > 15$\arcmin, following \citet{Boese 00}. We
evaluated the background in an annulus concentric to the source radius,
with radii depending on the off-axis. 
When the source was particularly faint, we calculated count rates in a
circle of radius corresponding to a smaller fraction of the PSF, to
increase the signal-to-noise ratio. We then corrected the count rate
accordingly, following \citet{Boese 00}.

We compared count rates obtained in this way with those reported in the 
BMW -- HRI catalogue and we found a general agreement, with the exception 
of some sources whose extension had been largely underestimated by the 
wavelet algorithm, as stated above. Indeed, our count rates would 
be the equivalent to the ``counted count rates'' reported in the catalogue, 
rather than those computed with the wavelet algorithm. In the comparison 
with this quantity we found a systematically higher count rate, 
consistent with the larger PHA interval used ($1 - 10$ in our 
analysis and $2 - 9$ in the BMW -- HRI catalogue).

The resulting net count rates are given in Table \ref{sampletab},
corrected for vignetting and lost counts due to the PSF (for point
sources only).

\subsection{Fluxes and luminosities} \label{fluxsec}
The count rates were converted into $0.1 - 2$ keV fluxes using a
conversion factor corresponding to a bremsstrahlung spectrum with kT =
5 keV plus the line of sight absorption appropriate for each source
from \citet{Dickey&Lockman 90} (reported in Table \ref{nh}). Although 
this spectrum might not be suitable for all kinds of
sources, the flux in the \textit{ROSAT} energy window depends only
weakly on the spectral model assumed, while it is more dependent on low
energy absorption. The resulting fluxes are in the range $10^{-14} -
10^{-11}$ \ergcmsec.

To calculate luminosities, we used distances listed in Table
\ref{sampletab}. The corresponding range in \lx is $10^{38} - 10^{43}$
\ergsec{}. 

\begin{table}
\begin{center}
\begin{tabular}{lc}
\hline \hline
$N_{\mbox{H}}$ (cm$^{-2}$) & CF (erg cm$^{-2}$ count$^{-1}$)\\
\hline \hline
$1 \times 10^{20}$ & $3.7 \times 10^{-11}$\\
$2 \times 10^{20}$ & $4.1 \times 10^{-11}$\\
$5 \times 10^{20}$ & $4.9 \times 10^{-11}$\\
$8 \times 10^{20}$ & $5.5 \times 10^{-11}$\\
$3 \times 10^{21}$ & $9.3 \times 10^{-11}$\\
\hline \hline
\end{tabular}
\end{center}
\caption{Galactic $N_{\mbox{H}}$ and corresponding unabsorbed flux in the band
0.1 - 2 keV for 1 count/s, assuming a  thermal bremsstrahlung spectral model with
kT = 5 keV.}
\label{nh}
\end{table}

\section{The atlas} \label{atlas}
\begin{figure*}[!h]
\centerline{
\psfig{figure=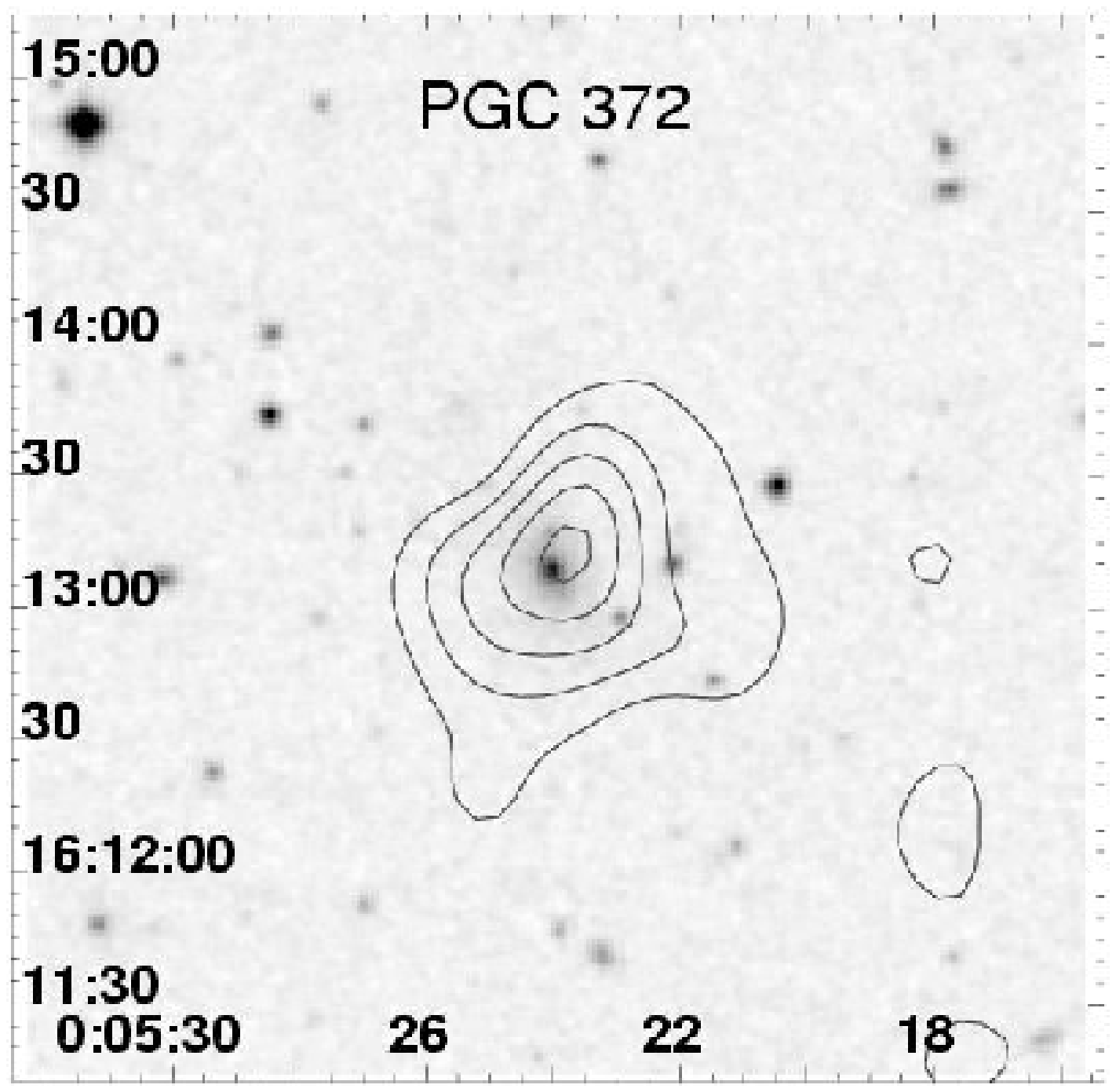,rheight=5cm,height=4.5cm,width=4.5cm,clip=}
\psfig{figure=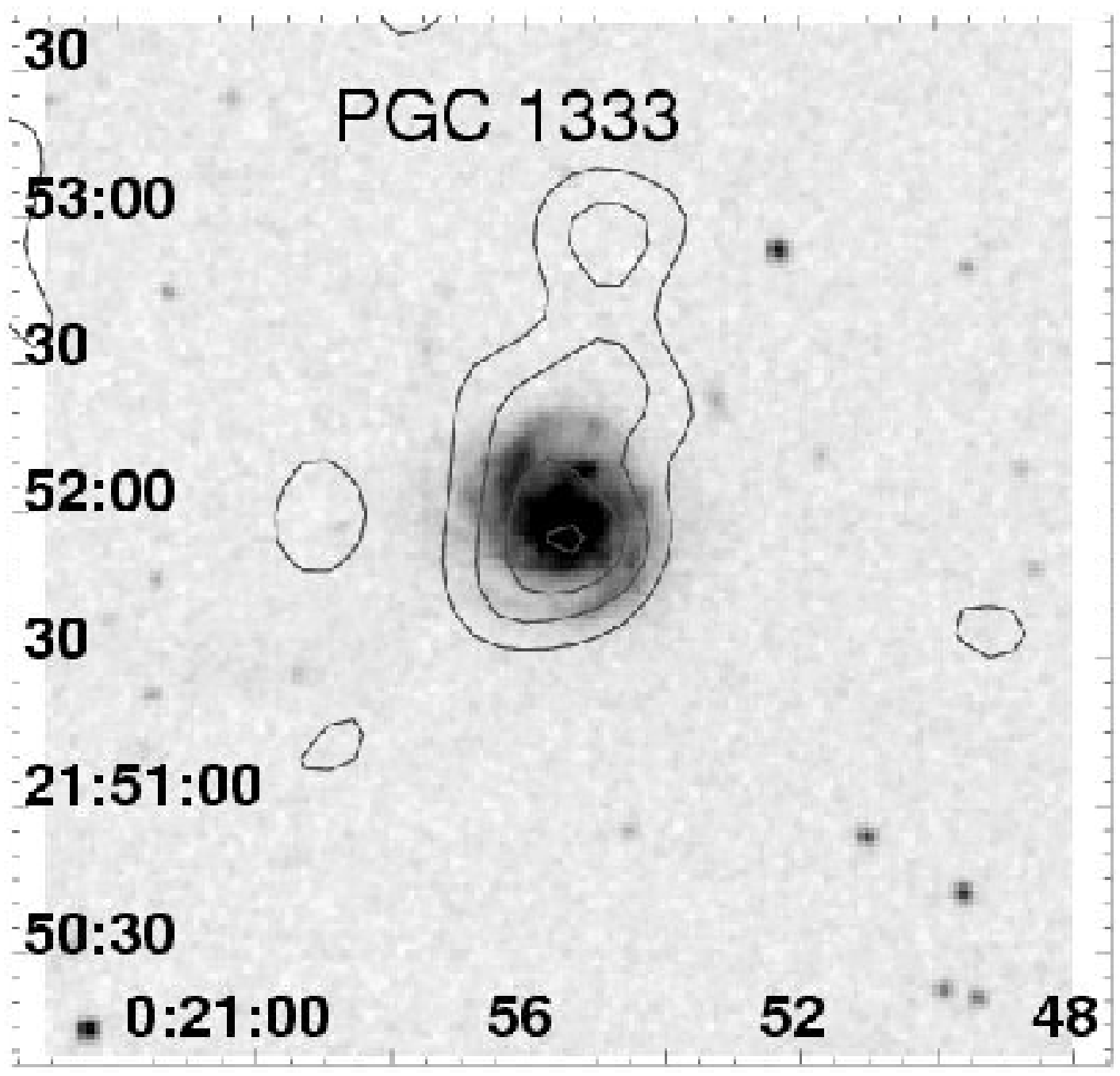,rheight=5cm,height=4.5cm,width=4.5cm,clip=}
\psfig{figure=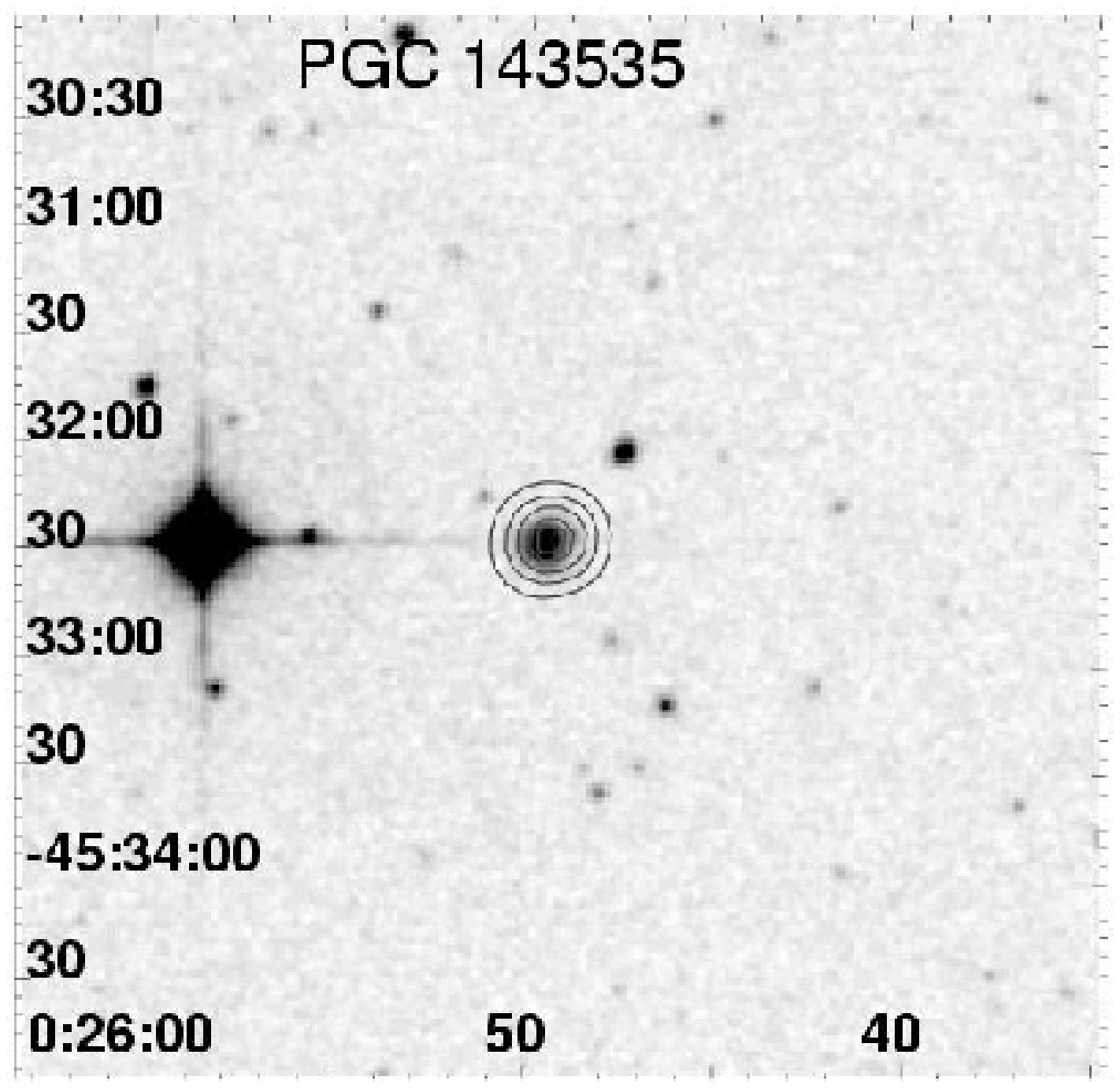,rheight=5cm,height=4.5cm,width=4.5cm,clip=}}
\caption{First page of the figure available in electronic form, that 
presents the \x contours from smoothed \x images superimposed onto optical images. 
Galaxies are generally at the center of the field. Smoothing is done with a 
Gaussian function of $\sigma$ = 5$''$ for point-like sources and of 
$\sigma$ = 10$''$ for extended sources.}
\label{atlasfig}
\end{figure*}

We provide in Fig. \ref{atlasfig} an overlay of \x contours of the
detected galaxies onto optical images from the Digital Sky Survey II
(DSS II) available from the ESO archive \footnote{\tt
http://archive.eso.org/dss/dss}. The Figure is available in electronic edition. 
We report here the first page as an example. When available, we used optical
images obtained with the blue filter, 
otherwise we used those obtained in the red filter. For PGC~209730 only
the DSS I plate is available.

Galaxies, ordered in RA and generally at the center of the field, 
are identified by their PGC name.

\x contours are produced from images in the PHA range $1-10$, smoothed with
a Gaussian function with $\sigma$ = 5\arcsec{} for point-like sources
and with $\sigma$ = 10\arcsec{} for extended sources.

\
\section{Optical observations} \label{optical}
In order to measure redshifts and magnitudes for some of the galaxies
of our sample, we made spectroscopical and photometrical observations at
the 1.52 meter telescope of the Osservatorio Astronomico di Bologna, at
Loiano (Italy) on the nights of the 16th and 17th October 2001. Because
of bad atmosferic conditions, we were able to observe only 3 galaxies.
A spectrophotometric calibration star was also observed. We present the
results obtained in Appendix \ref{opticalap}.

\section{Comparison with literature results} \label{luminosity}

\setcounter{figure}{2}%
\begin{figure}
\psfig{figure=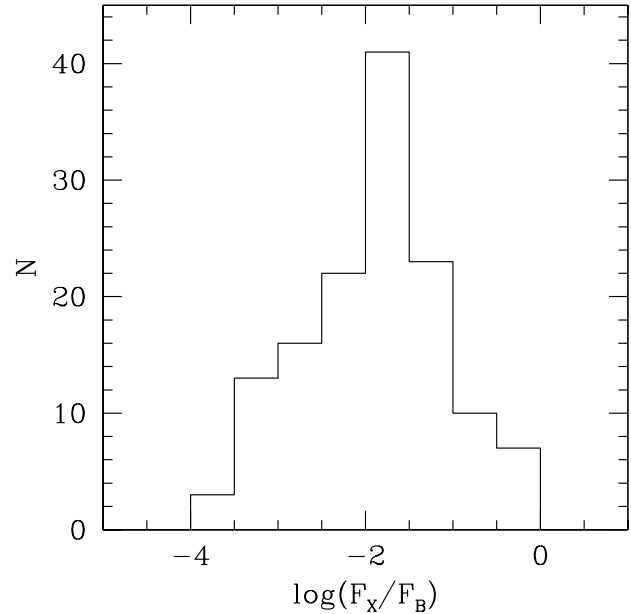,width=8.8cm,clip=}
\caption{\x - to - optical ratio distribution for the total sample. 
Both fluxes are in units of \ergcmsec.}
\label{xott}
\end{figure}

To verify whether the total sample of 143 objects is representative of
the \x properties of normal galaxies, we calculated \x luminosities
where possible (Table \ref{sampletab}) and we
plotted the distributions of \Lx, \Lb, the ratio \Lx/\lb and the
\Lx-\lb relationship for spiral and early-type galaxies. For the 19
galaxies for which the redshift is not known but we have m$_B$, the
ratio is calculated from fluxes. We compared our results with those in
the literature and found a good general agreement.

In particular:
\begin{enumerate}
\item the bulk of the galaxies in the sample has an \x luminosity
between 10$^{38}$ erg s$^{-1}$ and few 10$^{42}$ \ergsec, in accordance
with, e.g., \citet{Fabbiano 89}; 7 objects have \lx $\ga 10^{43}$
\ergsec, but most of them lie in a group, so that the intergalactic
medium could contribute to their luminosity; we cannot exclude the
presence of an unidentified AGN for some of these objects.

\item For spiral galaxies we found a linear relationship between
\x and optical luminosities (L$_X \propto$ L$_B^{1.0 \pm 0.2}$), 
in agreement with \citet{Fabbiano etal 92}. In a subsequent, 
more complete statistical analysis of the 234 ``normal'' 
spiral and irregular galaxies reported by \citet{Fabbiano etal 92}, 
\citet{Shapley etal 01} and \citet{Fabbiano&Shapley 02} find 
however that the \lx $-$ \lb relationship is significantly 
steeper than linear, with a slope of about 1.5. Our study 
is based on a much smaller sample (32 spiral and irregular 
galaxies), and small statistics could account for the discrepancy, 
as pointed out by \citet{Fabbiano&Shapley 02} in comparing their 
findings with previous works.

\item For early-type galaxies we found a steeper relationship, \Lx
$\propto$ \Lb$^{1.6 \pm 0.2}$, consistent with those obtained by
\citet{Fabbiano etal 92} and \citet{Eskridge etal 95}.

\item Values of the \x - to optical ratio in our sample cover roughly
the same range ($-4 <$ log(\Lx/\lb) $<$ 0 with \lx and \lb in \ergsec) 
as the spiral galaxies in \citet{Shapley etal 01} and
the early-type galaxies in \citet{Eskridge etal 95}, but their
distributions are different. We plot in Fig.~\ref{xott} the histogram
of the \x - to - optical ratio of the total sample of 143 galaxies. We
will analyze this subject in greater detail in subsection \ref{compxott}.

\end{enumerate}

\section{The logN-logS distribution} \label{lognlogs}

\input tablecompl.tex

The complete serendipitous sample of galaxies derived in section 
\ref{complete} (listed in
Table~\ref{compltab}) was used to calculate the integral flux distribution
(logN-logS) of normal galaxies with \x emission above the \x flux limit 
of the BMW -- HRI Catalogue
and B - magnitude $\leq$ 15.5.

\begin{figure}
\psfig{figure=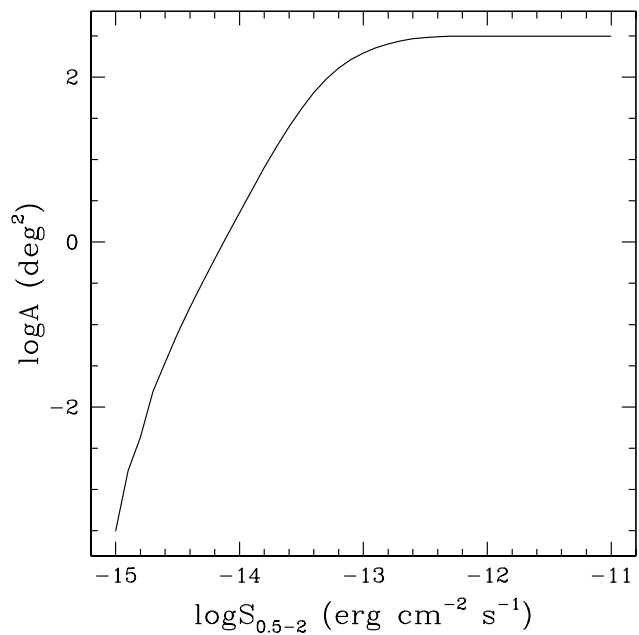,width=8.8cm,clip=}
\caption{Sky coverage for  the fields at galactic latitude $|b| 
\geq$ 10\degr {} and off-axis angle 3\arcmin $\leq \theta 
\leq 18$\arcmin, computed by assuming a thermal bremsstrahlung 
spectrum with kT = 5 keV and galactic line of sight absorption.}
\label{skycovfig}
\end{figure}

The sensitivity of the HRI instrument is not uniform over the entire
field of view. Moreover the observing time is different for different
fields so we need to calculate the area surveyed at any given flux (sky
coverage).

In the BMW -- HRI catalogue the published sky coverage was calculated by
means of simulations \citep[see][]{Panzera etal 03}. In this work we
used a sky coverage calculated with the same procedure, but with
parameters that reflect our source selection criteria. Therefore we 
included only fields with galactic
latitude $|b| \geq$ 10\degr, we considered only an annular region with
3\arcmin $\leq \theta \leq 18$\arcmin {} (the lower limit is to account
for the target region; the upper limit is the largest radius within 
the field of view of the detector in
the assumption of circular symmetry) and we assumed a bremsstrahlung
spectrum with kT = 5 keV plus the line of sight absorption.

The resulting sky coverage is plotted in Fig. \ref{skycovfig}. The
maximum area is $\sim$ 314 deg$^2$ and corresponds to fluxes above
$\sim 10^{-12}$ \ergcmsec. The surveyed area 
is $\sim$ 196 deg$^2$ at $10^{-13}$ \ergcmsec{} and $\sim$ 3 deg$^2$ 
at $\sim 1.1 \times 10^{-14}$ \ergcmsec{} (the lowest flux for the 
galaxies in our sample).

\begin{figure}
\psfig{figure=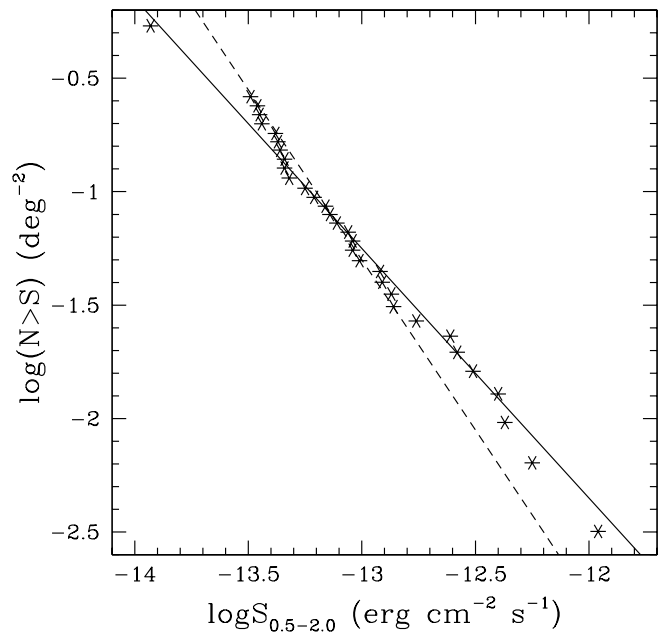,width=8.8cm,clip=}
\caption{The integral logN-logS distribution for the complete
serendipitous subsample (asterisks);
the solid line represents the $-1.1$
slope and the dashed line the euclidean $-1.5$ slope.}
\label{lognlogsfig}
\end{figure}

For consistency with the sky coverage calculation, the fluxes given in 
Table \ref{compltab} are derived from the original count rates estimated
in the BMW -- HRI catalogue using the wavelet algorithm and PHA channels
of HRI from 2 to 9, which are corrected for vignetting and PSF according to
\citet{Campana etal 99} and computed for the energy band $0.5 - 2$
keV, the standard of BMW -- HRI Catalogue.

The integral logN-logS distribution of the sample is shown in Fig.
\ref{lognlogsfig} and covers two decades in flux, from $\sim 1.1$ to
$\sim 110 \times 10^{-14}$ \ergcmsec. The overall distribution could be
approximated with a slope of $\sim -1.1$ (solid line in Fig.
\ref{lognlogsfig}). However the euclidean slope of $-1.5$ (dashed line)
is also consistent with the data: the excess of galaxies at the highest
fluxes is small and consistent within the limited statistics. Moreover,
at the lower fluxes, we could have some problems with incompleteness,
as we discuss below.

\subsection{Comparison with the literature: \x - to - optical ratio distribution} \label{compxott}

Before comparing our logN-logS distribution with those of other
samples in the literature, we need to better investigate the \x - to -
optical ratio distribution of our complete sample. 
We will consider samples 
derived from \textit{ROSAT} and \textit{Einstein} observations
that cover a flux range similar to ours. 
We will include samples derived from 
\textit{XMM - Newton} and
\textit{Chandra} surveys, that cover a flux range significantly
fainter than ours, in the discussion 
of the logN-logS
(subsection~\ref{complogn}).

The best available comparison could be with the sample of
\textit{candidate normal galaxies} found by \citet{Zimmermann etal 01}
in the \textit{ROSAT} All Sky Survey (RASS) and with the normal
galaxies found in the Einstein Extended Medium Sensitivity Survey
\citep[EMSS;][]{Gioia etal 90}, both \x selected. An effectively optically
selected sample for comparison is the \textit{Einstein} galaxy sample
\citep{Fabbiano etal 92, Shapley etal 01, Eskridge etal 95}.

\citet{Zimmermann etal 01} made a correlation study of the RASS Bright
Source Catalogue \citep{Voges etal 99} with the Catalogue of Principal
Galaxies \citep{Paturel etal 89}, 
from which they selected a sample of 198 \textit{candidate galaxies},
i.e. \x sources whose optical counterpart was not designated as AGN in
the literature. These selection criteria are similar to ours, and the
Catalogue of Principal Galaxies is a preliminary version of the current
LEDA database, so the two samples can be easily compared; however 
most of the
\citet{Zimmermann etal 01} sources have fluxes above $10^{-12}$
\ergcmsec{} (computed in the $0.1 - 2.4$ keV band, assuming a power law 
spectrum with photon index $\Gamma = 2.3$).

The EMSS was obtained from the analysis of 1453 images of the imaging
proportional counter (IPC) on board the \textit{Einstein} Observatory.
The survey covers an area of 778 deg$^2$ at $|b| > 20$\degr {} with
limiting sensitivity ranging from $\sim 5 \times 10^{-14}$ to $\sim 3 \times
10^{-12}$ \ergcmsec{} ($0.3 - 3.5$ keV band). 835 serendipitous sources 
were detected at or above the 4
$\sigma$ level \citep[see][]{Gioia etal 90, Stocke etal 91}. Among
these, 17 were identified as normal galaxies.

The \textit{Einstein} sample is the catalogue of normal galaxies observed 
by the \textit{Einstein} satellite, compiled by \citet{Fabbiano etal 92} 
and reanalyzed by \citet[ for early-type galaxies]{Eskridge etal 95} and 
by \citet[ for spiral galaxies]{Shapley etal 01}; we no longer distinguish 
here the two samples since early and late type galaxies are mixed in 
the BMW -- HRI sample.

\begin{figure}
\psfig{figure=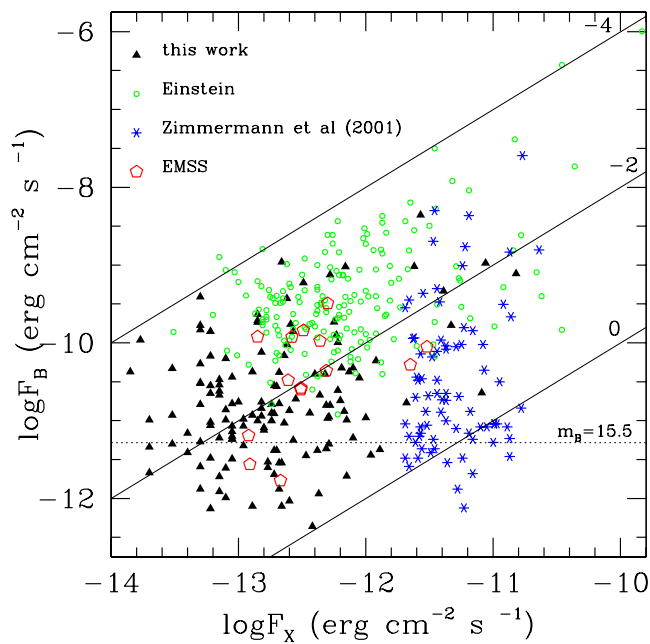,width=8.8cm,clip=}
\caption{B flux versus \x flux for our total sample
(solid triangles), for the \textit{Einstein}
sample (empty circles), for the Zimmermann et al. 2001 sample 
galaxies (asterisks), and for the EMSS galaxies 
(empty pentagons). Solid lines
correspond to log(F$_X$/F$_B$)= 0, $-2$, $-4$, as indicated; 
the horizontal dashed line is
the m$_B$ = 15.5 limit. We have considered only detections 
in the \textit{Einstein} sample: the exclusion of upper 
limits should not affect the results, since their distribution 
is consistent with that of the detections 
(see Fig. 5 in Shapley et al. 2001). We do not plot objects 
at F$_{0.1 - 2} < 2 \times 10^{-12}$ \ergcmsec{} in the 
Zimmermann et al. (2001) sample because below this flux 
their sample is not complete (see text).}
\label{fluxconf}
\end{figure}

In Fig. \ref{fluxconf} we plot the distribution of F$_X$ and F$_B$ values 
from all samples considered. For consistency with the values in Table~\ref{sampletab}, 
all fluxes are converted to the $0.1 - 2$ keV energy band, using a 
thermal bremsstrahlung spectrum with temperature kT = 5 keV and 
N$_H = 3 \times 10^{20}$ cm$^{-2}$.

Fig. \ref{fluxconf} clearly indicates that galaxies belonging to 
different samples populate different regions in the plot. Our sample 
(solid triangles) has an \x flux range between $10^{-14}$ and 
$10^{-11}$ \ergcmsec{} and log(F$_X/$F$_B$) between $-4$ and 0 
(see also Fig. \ref{xott}). The bulk of \textit{Einstein} galaxies 
(empty circles) is tipically at higher average fluxes and at lower 
values of log(F$_X$/F$_B)$ (between $-4$ and $-2$) compared to our 
distribution and the X-ray selected
samples in general.  This sample is the largest, it is effectively
optically selected and reasonably clean of the contamination from AGN
\citep{Shapley etal 01}. However, since it is not complete, it might 
not provide the true distribution of the \x - to - optical ratios.

The second largest sample is derived from \citet{Zimmermann etal 01}, 
with an additional flux limit F$_X > 2 \times 10^{-12}$ \ergcmsec{} 
(for completeness, see discussion below) 
and an upper limit in luminosity at logL$_X = 42.7$ \ergsec{} 
(the highest luminosity
in our complete sample), comparable to their limit to exclude potential
AGNs from the sample. The distribution of \textit{candidate galaxies} 
of \citet{Zimmermann etal 01} (asterisks) is significantly different from 
that of the \textit{Einstein} sample and extends at log(F$_X/$F$_B) > 0$.

The F$_X/$F$_B$ distribution for EMSS galaxies (empty pentagons) is at 
intermediate values and more consistent with that of our sample.

Also plotted in Fig. \ref{fluxconf} is the optical flux limit applied to 
our sample, m$_B = 15.5$. It is evident that the exclusion of galaxies 
fainter than this for the logN-logS calculation has an effect 
that increases as flux decreases. 
We have attepted to quantify it in order to correct the curve for lost objects.
Unfortunately we could not properly estimate the correction
because none of the
galaxy samples available allows us to derive the true distribution of \x
- to - optical ratios of normal galaxies. The EMSS should represent the
true F$_X/$F$_B$ distribution, but, given the small number of galaxies
(17, of which only 15 with B-magnitude), statistical errors are large.
The difference in the distribution of ratios between the two larger
samples in Fig.~\ref{fluxconf} suggests that they might be affected by
opposite biases: the \citet{Zimmermann etal 01} sample is likely to
contain unclassified AGNs, while the \textit{Einstein} sample could
lose objects at the highest \x - to - optical ratios. If we use the
three samples to estimate how many galaxies are lost as a function of
\x flux, we find that the corrections to the logN-logS are
small and the recomputed curve is consistent with the euclidean slope
in the observed flux range.

We also tried to estimate the correction by considering optically
fainter galaxies. At m$_B = 16$, LEDA is about 90\% complete \citep[see
Fig. 7 in][]{Paturel etal 97}: if we include galaxies down to this flux
limit, we only add 4 objects to our serendipitous sample, distributed
over the whole range of \x fluxes. Their inclusion only has a marginal
influence on the normalization, and none on the slope of the
distribution.

We conclude that, since the effects introduced by the 
optical limit are small, 
the logN-logS we derive is consistent with the euclidean slope.

\subsection{Comparison with the literature: logN-logS} \label{complogn}

\begin{figure}
\psfig{figure=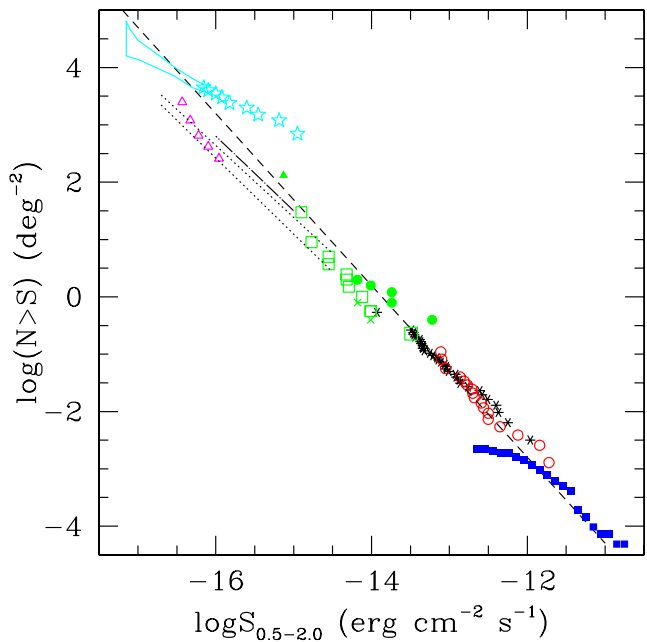,width=8.8cm,clip=}
\caption{Comparison of the logN-logs distribution obtained from
the present sample (asterisks) with several other relations for normal
galaxies from the literature.  The dashed line respresents the euclidean
relationship of Figure~\ref{lognlogsfig} extrapolated to the full flux range
considered. Symbols refer to: 
EMSS galaxies (empty circles);  \textit{candidate
galaxies} in Zimmermann et al. 2001 (solid squares); normal galaxies
(crosses), including LLAGN (solid circles) (Georgakakis et al. 2004a), 
and stacking analysis (solid
triangles, Georgakakis et al. 2003), found in the \textit{XMM-Newton}/2dF
survey; \textit{XMM-Newton}/Sloan DSS normal galaxies
from Georgakakis et al. (2004b) (empty
squares); \textit{Chandra} Deep Fields results from Hornschemeier et al. 
(2003) (empty triangles); optimistic or
pessimistic curves given by Bauer et al. (2004) (upper/lower dotted line);
Bayesian sample of Norman et al. (2004) reported by Ranalli et al. (2004)
(dot -- dashed line).  Open stars and horned-shaped box indicate the
detections and fluctuation analysis predictions of Miyaji \& Griffiths
(2002).
}
\label{lognlogsall}
\end{figure}

Figure \ref{lognlogsall} shows the comparison between the 
logN-logS derived above with several available from the literature.  
All fluxes are recomputed in the $0.5 - 2$ keV range that we use.

We find an excellent agreement with other samples that cover similar or
brighter flux ranges than the present sample. 

The \textit{candidate galaxies} \citep{Zimmermann etal 01}
appear to connect smoothly with
the euclidean extrapolation of the BMW -- HRI logN-logS above $10^{-12}$ \ergcmsec.
We interpret the flattening observed in the 
\citet{Zimmermann etal 01} data at lower fluxes 
as a result of their selection criteria. In fact, they indicate a $\ge$ 90\% 
completeness for count rates  $\ge$ 0.1 count s$^{-1}$, which converts to 
a flux of F$_{0.5-2} \sim 7 \times 10^{-13}$ \ergcmsec.

Although small, the EMSS is  
a truly complete sample, since it is
serendipitously X-ray selected and virtually completely identified
\citep{Gioia etal 90, Stocke etal 91, Maccacaro etal 94}. 
The EMSS appears to be euclidean and almost coincident with our curve 
for fluxes above $10^{-13}$ \ergcmsec.

We have also extended the comparison to include samples at at fainter fluxes. 
\citet{Georgakakis etal 04a} have computed the logN-logS of sources in
the XMM - Newton/2dF survey, obtained with the EPIC instrument on board the
\textit{XMM - Newton} satellite. This survey covers an area of about
2.5 deg$^2$ to the flux limit $\sim 10^{-14}$ \ergcmsec{} in the $0.5 -
8$ keV band (or F$_{0.5 - 2} \sim 5 \times 10^{-15}$ \ergcmsec). They
find two ``normal'' galaxies 
in their sample, that imply a density of $< 1$ source at F$_X  \sim
10^{-14}$ \ergcmsec{}, lower than what we find. 

However, the sample of \citet{Georgakakis etal 04a} contains three
additional galaxies (at $z \leq 0.1$ with L$_{0.5 - 8} \sim 10^{42}$
\ergsec) 
that the authors do not consider
because they might contain Low Luminosity AGNs (LLAGNs,see their Fig. 2).
Since we cannot exclude that our sample also contains a few  LLAGNs 
(see above and section \ref{luminosity}), 
we should consider these objects for a better
comparison with our sample. With the inclusion of these objects 
the  logN-logS better matches ours. 

Also shown in Fig. \ref{lognlogsall} are the
constraints from the stacking analysis results of \citet{Georgakakis
etal 03} at fainter fluxes, computed from optically selected galaxies 
at a mean redshift of $z \sim 0.1$.  The point derived from the 
total sample considered is
in excellent agreement with our logN-logS.

Recently \citet{Georgakakis etal 04b} have presented a pilot
sample of normal galaxies serendipitously detected in \textit{XMM - Newton}
public observations. They found 11 ``normal'' galaxy candidates 
with luminosities below $10^{42}$ erg s$^{-1}$ over an area of $\sim
4.5$ deg$^2$. They find that the LogN-LogS derived from this sample 
(plotted in Fig.
\ref{lognlogsall}) is again almost euclidean in slope, although at a 
smaller normalization than ours.

We also plot the results from deeper surveys, using  data from 
\citet{Hornschemeier etal 03}, \citet{Norman etal 04} as reported by 
\citet{Ranalli etal 04}, \citet{Bauer etal 04}, derived from the 
\textit{Chandra} Deep Fields. All these relations fall close to, although in general below, 
the extrapolation from our sample.  Different authors derive different
slopes for their samples, but they are all consistent with the
euclidean one. 

The lower normalizations found in these latter samples could be explained
in part by the combined effects of  more stringent criteria to minimize
contamination from the AGNs, even though of low luminosity, and of different
relative occurence of the galaxy types (spiral/starburst vs early
types).  

As already discussed, some residual contamination from low luminosity
AGNs could be present in the sample we have considered, since we  have
little information on the optical spectra, and could only reject known
AGNs.  We note however that \citet{Zimmermann etal 01} have applied the
same criterium adopted by \citet{Georgakakis etal 04a}, namely an X-ray
to optical luminosity ratio smaller than $10^{-2}$, and that the EMSS
sample, which is well studied optically, should not be contaminated by
AGNs. We have nevertheless considered discarding sources in our sample
that have a log(F$_{0.5-2}/$F$_B) < -2$.  We have 7 objects that
violate this limit, mostly at the high flux end. The resulting
logN-logS relation is slightly steeper, but consistent with that
presented in Fig.~\ref{lognlogsfig}, and would not significantly lower
the normalization of the ``euclidean" curve plotted.  However the
location of the different points from \citet{Georgakakis etal 04a} that
consider/discard possible contamination from LLAGNs give an idea of the
possible uncertainties involved.

The effect of different relative contributions from the early/late
types is more complicated to assess.  The lowest flux points
\citep{Hornschemeier etal 03} are derived from late type galaxies, so
they could underrepresent the total population.  However,  \citet{Bauer
etal 04} suggest that the early type galaxies follow a flatter
distribution, and in any case their LogN-logS is always below those
from starburst/quiescent galaxies (see their Fig. 9), so their
contribution could in fact be negligeable at the fainter fluxes.  In
the range covered by \citet{Georgakakis etal 04b}, there is only one
early type galaxy (but the sample is very small), while in our sample
we have a sizeable fraction of early types (1/3 among the objects with
a morphological classification), and the percentage increases to
$\sim$50\% in the ``candidate normal galaxies" of \citet{Zimmermann
etal 01}, for which moreover there is no apparent significant different
in the two slopes either.  The stacking analysis results from 
\citet{Georgakakis etal 03} derived separately for E/S0 and Sa-Scd 
bracket the extrapolation of the euclidean
logN-logS obtained from brighter samples, with Sa-Scd in better agreement
with the samples at low fluxes.
Since the emission from early and late type galaxies is due to
significantly different processes 
\citep{Fabbiano 89, Fabbiano etal 92, Kim etal 92, Eskridge etal 95, Shapley etal 01}, 
a different evolution is not out
of the question.  
An assessment of the local LogN-logS separately for the different
morphological types would be a step forward to a better understanding
of the properties of galaxies as a class, and would provide stronger
constraints for the investigation of normal galaxies at higher
redshifts. Investigating this aspect is beyond the scope 
and the potential of this work.  For the present time, we 
simply notice how remarkable it is that, in spite of the different
selection criteria and instruments used to define all the samples 
considered, the surface densities of normal galaxies is
consistent with a single euclidean
distribution for about 6 decades in flux (from $\sim 10^{-11}$ to $\sim
3 \times 10^{-17}$ \ergcmsec).

\section{Conclusions}

We  present the results for an ``almost serendipitous" sample of 143 \x
emitting normal galaxies selected  from the cross-correlation of the
BMW -- HRI Catalogue and the LEDA database.
Isointensity \x contours are overlayed onto the optical images for all 
galaxies and are presented in an atlas in Fig.~\ref{atlasfig}. 
The X-ray characteristics of the sample, listed in Table~\ref{sampletab},
are derived in a uniform way and are used in comparison with
other samples in the literature.  We find that  the
general properties of the total sample are in good agreement with those
already known for normal galaxies.

We have also selected a complete
subsample of 32 truly serendipitous sources in the local universe ($z < 0.07$), for which we
derived the logN-logS distribution in the flux range between $\sim 1.1$
and $110 \times 10^{-14}$ \ergcmsec, in the  $0.5 - 2$ keV
energy band. We find that the relation is consistent with the
euclidean distribution.

Moreover, we find a good agreement between our logN-logS and those
derived from \textit{ROSAT} PSPC and \textit{Einstein} data at 
similar or brighter
fluxes and from \textit{XMM - Newton} and \textit{Chandra} at fainter fluxes:
the overall distribution appears to be consistent with an euclidean
slope for about 6 decades in flux, from $\sim 3 \times 10^{-17}$ to
$10^{-11}$ \ergcmsec. The normalizations of different samples are
consistent within a factor of $\sim 2$.

Although with limited statistics, this work provides a first estimate
of the number density of sources identified with normal galaxies in the
nearby universe, in a flux range (both optical and X-ray) easily
accessible for follow up detailed observations.  This will
allow us to provide a solid basis that will help in studying
and classifying objects found in deeper surveys.

While current efforts are mainly focused on probing the distant
universe to determine the relevance of normal galaxies as a class at
very faint fluxes, the success of these studies depends also on
the constraints given by the  bright flux end of the
number counts, which is still poorly studied. This is particularly relevant at the lowest fluxes where the number counts
approach the total source density (e.g. the detections and fluctuation
analysis results of \citet{Miyaji&Griffiths 02}, see 
\citet{Hornschemeier etal 02} and Fig. 7). 
The sample here derived will be therefore instrumental
to studies of both the
galaxy cosmological evolution and the contribution of this class of
sources to the X-ray background.

\begin{acknowledgements}
We thank Dr. R. Panzera for her invaluable help, in particular 
in the extraction of our sample from the BMW -- HRI catalogue.\\
We thank Dr. U. Zimmermann for useful discussion on their work 
and for providing us with the original data for their logN-logS 
that we use in Fig. \ref{lognlogsall}.\\
We acknowledge partial financial support from the Italian Space Agency (ASI).\\
This research has made use of the Lyon-Meudon Extragalactic Database (LEDA), 
{\tt http://www-obs.univ-lyon1.fr/hypercat/} or 
{\tt http://www.brera.mi.astro.it/hypercat/}.\\
This research has made use of the NASA/IPAC Extragalactic Database (NED) 
which is operated by the Jet Propulsion Laboratory, California Institute of 
Technology, under contract with the National Aeronautics and Space Administration.\\ 
This research has made use of the SAOimage ds9 developed by 
the Smithsonian Astrophysical Observatory.\\
The compressed files of the ``Palomar Observatory - Space Telescope 
Science Institute Digital Sky Survey" of the northern sky, based on
scans of the Second Palomar Sky Survey are copyright (c) 1993-2000 
by the California Institute of Technology and are distributed 
by agreement. All Rights Reserved.
\end{acknowledgements}

\input ref.tex
\clearpage

\appendix
\section{Results of optical observations} \label{opticalap}
The log of observations made at the Loiano Telescope in order to measure
 redshifts and magnitudes for some galaxies in our sample 
 (see section \ref{optical}) is listed in Table \ref{log}.

\begin{table*}
\begin{center}
\begin{tabular}{lccccc}
\hline \hline
Name &  \multicolumn{2}{c}{LEDA coordinates} & Spectroscopy & \multicolumn{2}{c}{Photometry}\\
 & R. A. (2000) & Dec. (2000) & Exposure (s) & Filter & Exposure (s) \\
\hline \hline
PGC 65435 & 20 48 43.09 & 25 16 53.0 & 2 $\times$ 1800 & R & 2 $\times$ 120 \\
PGC 69291 & 22 36 33.58 & 34 30 04.5 & 2 $\times$ 1800 & R & 1 $\times$ 60  \\
	  &	        &	     &		       & R & 3 $\times$ 40  \\
	  &		&	     &		       & V & 1 $\times$ 40  \\
	  &		&	     &		       & V & 3 $\times$ 80  \\
	  &		&	     &		       & B & 3 $\times$ 200 \\
PGC 196693 & 23 48 18.07 & 01 06 17.3 & 1 $\times$ 1800 &	& \\
Feige 110 & 23 19 58.40 & -05 09 56.2& 2 $\times$ 480  & R & 2 $\times$ 30 \\
	  &		&	     &		       & V & 2 $\times$ 30 \\
	  &	        &	     &		       & B & 2 $\times$ 40 \\
\hline \hline
\end{tabular}
\end{center}
\caption{Log of optical observations.}
\label{log}
\end{table*}

\begin{table*}
\begin{center}
\begin{tabular}{lc}
\hline \hline
Name & Redshift \\
\hline \hline
PGC 65435  & 0.0485 $\pm$ 0.0006 \\
PGC 69291  & 0.0290 $\pm$ 0.0009 \\
PGC 196693 & 0.093 $\pm$ 0.001 \\
\hline \hline
\end{tabular}
\end{center}
\caption{Average value of redshifts measured for the three galaxies observed.}
\label{redshift}
\end{table*}

The seeing was about 2\arcsec, so we used a long slit of width
2.5\arcsec{} for spectra. The spectral range was 3800 \AA{} - 8700 \AA.

For photometry we used the B - V - R filters of Johnson - Kron - Cousin system. 

We made use of the Image Reduction and Analysis Facility (IRAF)\footnote{IRAF is 
a general purpose software system for the reduction and analysis 
of astronomical data. IRAF is written and supported by the IRAF 
programming group at the National Optical Astronomy Observatories (NOAO) 
in Tucson, Arizona. NOAO is operated by the  Association of Universities 
for Research in Astronomy (AURA), Inc. under cooperative agreement with 
the National Science Foundation.}
for spectroscopic and photometric data reduction.

We used all lines visible in the spectra to measure $z$, which is
calculated from the average value. The redshifts obtained are
reported in Table \ref{redshift}. Only absorption lines, typical of 
early-type galaxies, and no sign of nuclear activity were detected 
in the galaxy spectra.

The atmosferic conditions were inadequate for photometric observations,
however we list our raw estimate of magnitudes in Table \ref{magnitude}
and we use the value obtained for PGC~69291 in Table \ref{sampletab},
where it is reported in brackets. 

\textbf{PGC 196693}\\
After obtaining the spectrum, we discovered that NED lists the
galaxy SDSS~J234817.99+010617.0 at the same position and with
the same redshift we obtained, although it does not give a PGC name. We
therefore identify PGC~196693 as SDSS~J234817.99+010617.0.

\textbf{PGC 69291}\\
The galaxy is identified as NGC~7325 by LEDA and as NGC~7327 by NED,
with a note suggesting that it could be a star. Our observations show 2
objects, a star and a galaxy at $z = 0.0290 \pm 0.0009$ at the position
of PGC~69291. Most of the optical flux is due to the star, as listed in
Table \ref{magnitude}.

The uncertainty in the \x position (see also the map in Fig. \ref{atlasfig}) 
does not allow us to identify the \x source with either object. Unfortunately 
the bad quality of our optical does not allow us to use the \x - to - optical 
ratio \citep[see for example][]{Maccacaro etal 88} to discriminate between them, 
because it is consistent with both the star and the galaxy. We list the galaxy 
in the total
sample but due to its faint estimated magnitude it is not included in the
complete serendipitous sample.

\begin{table*}
\begin{center}
\begin{tabular}{lccc}
\hline \hline 
Name & \multicolumn{3}{c}{Magnitude} \\
     & B & V & R \\
\hline \hline 
PGC 65435 & & & 14.8 \\
PGC 69291 galaxy & 16.6 & 16.1 & 15.6 \\
PGC 69291 star & 13.0 & 12.5 & 12.2 \\
\hline \hline
\end{tabular}
\end{center}
\caption{Estimated magnitudes for two of the three galaxies observed and for
the star found near PGC 69291 resulting from our observations with the Loiano
Telescope.}
\label{magnitude}
\end{table*}


\section{Table of total sample}
We report here the complete table with our total sample that will be available
in the electronic form. For explanation of the columns refer to the sample
Table~\ref{sampletab} in the main text.

\onecolumn
  \fontsize{8pt}{10pt}
  \begin{landscape}
  \begin{verbatim}
(1)                (2)         (3)        (4) (5) (6) (7)  (8)         (9)                           (10)  (11) (12)    (13)  (14)   (15)  (16) (17)  (18)
----------------------------------------------------------------------------------------------------------------------------------------------------------------   
BMW000523.9+161307 00 05 23.83 +16 13 11.3 e  52           PGC 0000372                                     698          15.74  2.49   7.40  1.17 43.59 Cl    
BMW002055.1+215208 00 20 55.43 +21 51 52.2 p               PGC 0001333 IC 1543                       Sbc   112  14.54    7.63  1.11   3.58  0.52 41.71 c     
BMW002055.2+215208                        		   	        		      	     	              	        
BMW002549.3-453227 00 25 49.26 -45 32 26.5 p               PGC 0143535                               Sab        16.57   18.64  3.16   7.64  1.30             
BMW002950.2-405630 00 29 50.99 -40 56 37.9 p               PGC 0130966 DUKST 294-9                   Sc    241  15.35   19.06  2.31   7.81  0.95 42.72 c     
BMW003652.3-333310 00 36 52.82 -33 33 14.7 p               PGC 0002204 ESO 350-IG38                  S?    123  14.96    2.74  0.43   1.12  0.14 41.27 c
BMW003918.5+030220 00 39 18.47 +03 02 14.8 p               PGC 0002362 NGC 194                       E     103  13.09   14.75  1.94   6.49  0.85 41.89 c; Gr
BMW003948.3+032219 00 39 48.77 +03 22 21.0 p               PGC 0002401 UM 57                                             6.15  0.49   2.70  0.22         
BMW004242.0+405154 00 42 42.48 +40 51 52.6 e  60           PGC 0002555 NGC 221                       E       1   8.18   52.44  1.06  26.75  0.54 38.52 Gr
BMW010716.2+323117 01 07 16.20 +32 31 16.6 p               PGC 0003966 NGC 379                       S0    110  13.48    1.37  0.35   0.67  0.17 40.98 Gr
BMW010717.4+322857 01 07 17.76 +32 28 57.0 p               PGC 0003969 NGC 380                       E      88  13.24    4.19  0.50   2.06  0.24 41.26 Gr
BMW010717.8+322857 	                  		   	        		      	     	              	        
BMW010918.1+131013 01 09 18.63 +13 10 11.4 e  60           PGC 0004117 UGC 716                             356  14.78    3.77  0.67   1.77  0.32 42.41 c; Cl
BMW011314.7+153059 01 13 15.84 +15 31 05.0 p               PGC 0004392                                     276           1.60  0.24   0.75  0.12 41.78 Cl 
BMW011315.3+153102                        		   	        		      	     	              	        
BMW011315.8+153059                        		   	        		      	     	              	        
BMW012038.2+294157 01 20 38.06 +29 41 56.2 p               PGC 0004848 IC 1672                       Sb    142  13.28    7.83  2.65   4.00  1.35 41.96 c     
BMW012038.8+294150                        		   	        		      	     	              	        
BMW012310.6+332738 01 23 11.33 +33 27 29.5 e  50           PGC 0005060 NGC 499                       E-S0   87  12.86   28.37  1.51  13.90  0.74 42.01 Gr    
BMW012311.5+332741                        		   	        		      	     	              	        		          
BMW012558.8-012023 01 25 59.10 -01 20 22.8 e  18           PGC 0005323 NGC 545                       E-S0  107  12.78    1.72  0.28   0.81  0.13 41.02 c; Cl
BMW012600.2-012040 01 26 00.44 -01 20 40.3 e  50           PGC 0005324 NGC 547                       E     110  12.68    4.83  0.69   2.27  0.32 41.50 c; Cl
BMW012600.4-012039                        		   	        		      	     	              	        		          
BMW012600.7-012040                        		   	        		      	     	              	        		         
BMW012623.1+013921 01 26 23.24 +01 39 17.7 p               PGC 0144123 APMUKS(BJ)B012348.87+012346.1            17.13    9.92  0.84   4.37  0.37         
BMW014358.8+022053 01 43 58.82 +02 20 52.5 e  40           PGC 0006367 UGC 1214                      S0-a  103  14.49   15.22  0.77   6.70  0.34 41.90 c
BMW015641.6+330238 01 56 41.61 +33 02 30.9 p               PGC 0007289 NGC 736                       E      87  12.72    1.17  0.33   0.60  0.17 40.71 c; Gr
BMW021853.5+423440 02 18 53.93 +42 34 38.9 p               PGC 0165335                               E     110           0.48  0.14   0.26  0.08 40.53 Cl
BMW022537.8+365759 02 25 37.90 +36 57 48.7 p               PGC 0009215 UGC 1877                      E     216  14.14   12.86  1.02   6.30  0.50 42.52 Cl    
BMW022602.5-232136 02 26 02.54 -23 21 36.0 p               PGC 0133905 DUKST 479-24                  Sc    337  15.59    2.79  0.07   1.14  0.03 42.15
BMW023741.8+015827 02 37 41.49 +01 58 27.5 p               PGC 0009961 NGC 1004                      E     130  13.70    5.19  1.18   1.84  0.52 41.58 Cl    
BMW023819.3+020706 02 38 19.35 +02 07 12.5 p               PGC 0009997 NGC 1016                      E     132  12.59    3.58  1.16   1.58  0.51 41.47 Cl    
BMW024539.7-525748 02 45 39.70 -52 57 47.5 e  40           PGC 0010446 ESO-LV 1540110                S0         15.46    1.11  0.40   0.48  0.18       c 
BMW031858.7+412808 03 18 58.42 +41 28 08.3 p               PGC 0012350 NGC 1270                      E      98  13.40    1.61  0.31   0.93  0.18 40.99 Cl
BMW032244.0-370614 03 22 44.13 -37 06 12.2 p               PGC 0012653 NGC 1317                      SBa    25  11.58    3.27  0.36   1.34  0.15 40.00 Cl
BMW032244.1-370613                        		   	        		      	     	              	        
BMW033051.3-523034 03 30 51.46 -52 30 36.3 p               PGC 0013046                               E-S0  345  14.98    2.21  0.89   0.87  0.36 42.08 Cl
BMW033657.5-353015 03 36 57.51 -35 30 15.1 p               PGC 0013344 NGC 1387                      E-S0   25  11.69    6.76  0.59   2.50  0.22 40.26 Cl 
BMW033728.0-243003 03 37 28.17 -24 30 05.8 p               PGC 0013368 NGC 1385                      SBc    26  10.87    6.20  1.59   2.29  0.59 40.26 c     
BMW033828.6-352659 03 38 28.90 -35 27 02.1 e 500           PGC 0013418 NGC 1399                      E      25  10.06  411.10 22.67 152.10  8.39 42.03 Cl    
BMW033828.9-352659                        		   	        		      	     	              	        
BMW033830.0-352711                        		   	        		      	     	              	        
BMW033851.5-353536 03 38 51.92 -35 35 41.9 e 150           PGC 0013433 NGC 1404                      E      25  10.62  111.00  1.21  41.06  0.45 41.46 Cl
BMW033851.6-353536                        		   	        		      	     	              	        
BMW033851.7-353536                        		   	        		      	     	              	        
BMW041205.3-325222 04 12 03.09 -32 52 28.9 e        75 150 PGC 0014638 NGC 1532                      SBb    20   9.69    5.16  1.26   2.12  0.52 40.00 Gr    
BMW041544.5-553533 04 15 45.64 -55 35 33.2 e  60           PGC 0014757 NGC 1549                      E      20  10.36    7.72  0.78   3.17  0.32 40.16 Gr
BMW041545.1-553537                        		   	        		      	     	              	        
BMW041545.8-553533                        		   	        		      	     	              	        
BMW042900.6-533653 04 29 00.91 -53 36 45.0 p               PGC 0015231 IC 2081                       E-SO  255  14.62    3.39  0.60   1.25  0.22 41.98 Cl 
BMW042900.7-533647                        		   	        		      	     	              	        
BMW043754.0-425852 04 37 55.01 -42 58 53.0 p               PGC 0130376                               E          15.53   12.38  2.04   5.08  0.84             
BMW044707.0-202847 04 47 06.29 -20 28 39.2 p               PGC 0099957                               E     430  15.41    6.05  1.63   2.66  0.72 42.75 Cl    
BMW050138.5-041527 05 01 38.26 -04 15 33.5 p               PGC 0016574 NGC 1741                      Sm     81  14.56    3.60  0.91   1.84  0.46 41.15 c; Gr 
BMW050943.5-083652 05 09 44.01 -08 36 51.9 p               PGC 0147040                               S0-a       17.00    0.89  0.34   0.49  0.19         
BMW053359.9-714524 05 34 00.41 -71 45 28.5 p               PGC 0017451 ESO 56-G154                         145  13.42   16.17  1.94   8.90  1.07 42.33 c     
BMW062457.1+821914 06 24 57.12 +82 19 13.8 p               PGC 0018991 UGC 3435                             86  14.19    3.17  0.82   1.55  0.40 41.10 c 
BMW062545.7-042202 06 25 45.47 -04 22 04.4 p               PGC 0075751 CGMW 1-0108                   Sc                 12.25  0.66  11.39  0.62          
BMW062545.8-042203                        		   	        		      	     	              	        
BMW071957.5-320041 07 19 58.78 -32 00 54.4 p               PGC 0077028 CGMW 2-0778                   Sbc        13.74   17.30  3.91  10.03  2.27             
BMW073217.8+854234 07 32 17.80 +85 42 34.1 e  22.5         PGC 0021231 NGC 2300                      E-S0   47  11.61   10.90  0.43   5.56  0.22 41.15 Gr 
BMW073220.2+854230 	                  		   	        		      	     	              	        
BMW074914.6-055607 07 49 14.56 -05 56 11.3 p               PGC 0078533 CGMW 1-1742                   Sc                  1.25  0.26   0.72  0.15         
BMW081023.2+421628 08 10 23.16 +42 16 27.1 e  66           PGC 0022928 CGCG 207-40                         379  15.90   22.15  1.17  10.85  0.57 43.25 Cl    
BMW082133.8+470242 08 21 33.55 +47 02 37.7 p               PGC 0023450                                     775  17.61    1.27  0.59   0.60  0.28 42.61 Cl 
BMW090825.4-093349 09 08 25.49 -09 33 41.9 p               PGC 0025701                                     958  17.27    1.69  0.26   0.82  0.13 42.92 Cl 
BMW090825.7-093336                        		   	        		      	     	              	        
BMW090851.5+110122 09 08 51.50 +11 01 22.0 e  30           PGC 0025741                                     978  17.07    1.50  0.31   0.66  0.14 42.88 Cl 
BMW090852.1+110138                        		   	        			      	              	        
BMW090943.9+071032 09 09 43.87 +07 10 31.5 p               PGC 0025825 NGC 2773                      Sb    109  14.26    1.05  0.32   0.50  0.15 40.83 c 
BMW091041.7+071223 09 10 41.86 +07 12 27.0 p               PGC 0025876 NGC 2777                      Sab    26  14.45    1.67  0.31   0.79  0.15 39.79 Gr 
BMW091826.3-122222 09 18 27.00 -12 22 26.9 e  75           PGC 0087445                                     315  15.68    9.97  1.13   4.89  0.56 42.74 Cl    
BMW100108.4-192631 10 01 08.92 -19 26 26.8 p               PGC 0028995 ESO 567-G3                    Sab        14.30   10.91  1.26   5.34  0.61       c; Gr?
BMW100108.7-192621                        		   	        		      	     	              	        
BMW100114.9+554316 10 01 15.17 +55 43 06.0 p               PGC 0028990 MCG 9-17-9                           31  15.20    1.10  0.38   0.41  0.14 39.64 c; Gr
BMW100115.1+554318                        		   	        		      	     	              	        
BMW100115.5+554308                        		   	        		      	     	              	        
BMW100158.3+554100 10 01 58.60 +55 40 49.0 e 120           PGC 0029050 NGC 3079                      SBcd   31   9.84   18.44  1.18   6.82  0.44 40.87 c; Gr 
BMW100158.3+554060                        		   	        		      	     	              	        
BMW100158.4+554103                        		   	        		      	     	              	        
BMW100158.7+554051                        		   	        		      	     	              	        
BMW101010.1+541828 10 10 10.10 +54 18 27.7 p               PGC 0029594 CGCG 266-27                         261  15.38    2.57  1.27   0.95  0.47 41.84       
BMW101825.0+215348 10 18 25.03 +21 53 45.9 e  19           PGC 0030099 NGC 3193                      E      35  11.72    1.21  0.31   0.50  0.13 39.84 Gr
BMW103633.4-271318 10 36 33.36 -27 13 17.4 e 106           PGC 0031453                                     215  14.71    9.12  2.81   4.47  1.38 42.42 Cl    
BMW103719.0-271121 10 37 18.92 -27 11 24.8 e  25           PGC 0031540 NGC 3315                      E-S0   75  13.84    1.39  0.36   0.68  0.18 40.65 Cl 
BMW104827.8+123157 10 48 27.76 +12 31 56.6 e  50           PGC 0032306 NGC 3389                      Sc     34  11.63    3.23  0.89   1.42  0.39 40.26 Gr
BMW104925.3+324639 10 49 25.13 +32 46 27.1 p               PGC 0032390 IC 2604                       SBm    50  14.18    2.45  0.51   1.01  0.21 40.45
BMW110522.3+381421 11 05 23.18 +38 14 13.2 p               PGC 0033514 CGCG 213-27                         170  15.92    9.06  2.70   3.35  1.00 42.05       
BMW113446.5+485718 11 34 46.51 +48 57 17.6 p               PGC 0035780 IC 711                        E-SO  194  14.95    1.39  0.41   0.57  0.17 41.41 Cl
BMW113448.9+490440 11 34 49.51 +49 04 39.6 p               PGC 0035785 IC 712                        E     201  13.67    1.48  0.45   0.61  0.18 41.44 Cl
BMW114223.6+101552 11 42 23.79 +10 15 51.2 p               PGC 0036348 NGC 3825                      SBa   128  13.57    1.26  0.33   0.55  0.15 40.97 Gr 
BMW114232.1+583706 11 42 32.04 +58 37 05.8 p               PGC 0036369 MCG +10-17-46                            15.82    1.21  0.32   0.45  0.12         
BMW114430.3+305305 11 44 30.20 +30 53 02.9 p               PGC 0139685                                     458  16.51    4.29  1.43   1.76  0.59 42.65 Cl    
BMW115535.8+232724 11 55 35.92 +23 27 27.5 p               PGC 0139725                                     815  17.55    3.24  0.57   1.33  0.23 42.99
BMW120438.2+014717 12 04 38.50 +01 47 10.5 p               PGC 0038218 NGC 4077                      S0    140  13.97    1.68  0.64   0.69  0.26 41.19 Gr
BMW120631.5-295119 12 06 31.46 -29 51 18.7 p               PGC 0157406                                          14.62    0.52  0.17   0.25  0.08       c
BMW121049.5+392822 12 10 49.82 +39 28 23.9 p               PGC 0038773 NGC 4156                      SBb   135  13.65    1.71  0.13   0.70  0.05 41.17 c
BMW121049.5+392820                        		   	        		      	     	              	        
BMW121049.6+392823                        		   	        		      	     	              	        
BMW121633.3+072746 12 16 34.07 +07 27 46.9 p               PGC 0039328 NGC 4224                      Sa     53  12.35    5.79  1.06   2.37  0.43 40.88 Cl    
BMW121908.2+470531 12 19 08.40 +47 05 20.5 p               PGC 0039615 UGC 7356                      Irr     8  15.35    1.87  0.47   0.96  0.17 38.65  
BMW122018.0+752214 12 20 17.39 +75 22 16.6 e  40           PGC 0039791 NGC 4291                      E      44  12.04   13.65  1.37   6.01  0.60 41.11       
BMW122503.5+125317 12 25 04.23 +12 53 07.5 e 100           PGC 0040455 NGC 4374                      E      25   9.84   54.48  2.77  23.97  1.22 41.23 Cl    
BMW122504.1+125308                        		   	        		      	     	              	        
BMW122504.3+125318                        		   	        		      	     	              	        
BMW122504.4+125316                        		   	        		      	     	              	        
BMW122504.4+125319                        		   	        		      	      	              	        
BMW122611.1+125642 12 26 11.68 +12 56 45.2 e       200 400 PGC 0040653 NGC 4406                      E      25   9.73  199.80  7.89  87.91  3.47 41.79 Cl    
BMW122612.6+125628                        		   	        		      	     	              	        
BMW122716.2+085815 12 27 14.97 +08 58 17.3 e  80           PGC 0094216                                     530  16.11    3.74  1.03   1.53  0.42 42.73 Cl    
BMW123017.6+121936 12 30 17.59 +12 19 35.7 e  30           PGC 0041297 NGC 4478                      E      25  11.93    1.31  0.30   0.68  0.13 39.63 Cl 
BMW123018.1+121926                        		   	        		      	     	              	        
BMW123018.1+121926                        		   	        		      	     	              	        
BMW123052.5+143307 12 30 52.51 +14 33 06.7 e  40           PGC 0169455 VPC 774                                  18.18    7.98  1.43   3.51  0.63             
BMW123643.2+131515 12 36 43.14 +13 15 19.0 p               PGC 0042081 IC 3583                       Irr    25  13.11    0.67  0.32   0.30  0.14 39.33 Cl
BMW124342.7+162340 12 43 42.68 +16 23 38.0 p               PGC 0042833 NGC 4651                      Sc     25  10.81    1.30  0.36   0.53  0.15 39.55 c; Cl
BMW125232.8-153057 12 52 33.03 -15 31 02.2 p               PGC 0043675 MCG -2-33-38                  SO     87  14.39    3.19  0.31   1.50  0.14 41.11 c; Gr
BMW125233.1-153102                        		   	        		      	     	              	        
BMW125302.3-152120 12 53 02.26 -15 21 19.7 p               PGC 0043748 DRCG25-68                     E-SO  266  15.55    0.54  0.20   0.25  0.09 41.38 Cl 
BMW125318.1-153203 12 53 18.27 -15 32 03.4 e  30           PGC 0043777 DRCG25-39                     E-SO  284  14.78    2.94  0.34   1.38  0.16 42.11 Cl
BMW125318.3-153203                        		   	        		      	     	              	        
BMW125749.0-171628 12 57 49.85 -17 16 25.3 e  50           PGC 0044369                               E     296  14.77   13.80  1.77   6.76  0.87 42.84 Cl    
BMW125750.2-171616                        		   	        		      	     	              	        
BMW130055.7+274740 13 00 55.99 +27 47 29.8 p               PGC 0044840 NGC 4911                      SBbc  159  13.30   12.83  1.03   4.74  0.38 42.14 Cl    
BMW130056.0+274737                        		   	        		      	     	              	        
BMW130056.2+274731                        		   	        		      	     	              	        
BMW130559.7+291647 13 05 59.46 +29 16 46.8 p               PGC 0045318 CGCG 160-136                  E     159  15.08    4.42  0.95   1.64  0.35 41.66 c 
BMW130803.0+292918 13 08 03.09 +29 29 12.0 p               PGC 0166974                                                   2.61  0.17   0.97  0.06       Cl?
BMW130803.2+292910                        		   	        		      	     	              	        
BMW130803.3+292911                        		   	        		      	     	              	        
BMW130803.4+292903                        		   	        		      	     	              	        
BMW131324.5-325247 13 13 24.96 -32 52 44.7 p               PGC 0158561                                          16.26    1.12  0.41   0.55  0.20           
BMW131413.5-425739 13 14 13.44 -42 57 38.8 e  60           PGC 0046023 NGC 5026                      SBa    75  11.73   86.49  3.71  47.57  2.04 42.48 Cl?   
BMW131912.4-123228 13 19 12.40 -12 32 28.1 e  20           PGC 0046432 NGC 5072                      E-S0  136  14.23    2.09  0.74   0.92  0.33 41.27 c
BMW132040.2-434200 13 20 40.20 -43 42 00.3 p               PGC 0046566 NGC 5082                      S0     78  12.96    0.93  0.50   0.51  0.27 40.54 Gr
BMW132745.6-314751 13 27 45.41 -31 47 54.0 p               PGC 0047177                                     265  14.53    3.74  0.61   1.76  0.29 42.15 Cl
BMW132759.6-272104 13 27 59.87 -27 21 29.6 p               PGC 0047209 IC 4255                       E     203  13.66    3.91  1.11   1.92  0.55 41.95 Cl    
BMW132802.6-314519 13 28 02.56 -31 45 19.0 p               PGC 0088857                               E     258  15.01    3.37  0.60   1.58  0.28 42.08 Cl 
BMW132958.2+471613 13 30 00.20 +47 15 55.1 e  80           PGC 0047413 NGC 5195                      SBpec  14  10.10   12.66  0.77   5.19  0.32 40.06 Gr
BMW132959.1+471555                        		   	        		      	      	              	        
BMW132959.8+471556                        		   	        			      	              	        
BMW133000.1+471558                        		   	        			      	              	        
BMW133006.4-014316 13 30 06.34 -01 43 06.8 e  45           PGC 0047432 NGC 5183                      Sb     86  13.21    2.86  0.98   1.17  0.40 41.04 c     
BMW133425.1+344127 13 34 24.86 +34 41 26.9 e  60           PGC 0047822 NGC 5223                      E     144  13.22   20.33  1.70   7.52  0.63 42.24 Cl    
BMW134623.6-375810 13 46 23.36 -37 58 23.3 p               PGC 0048817 ESO 325-G16                         228  14.67   11.43  1.27   5.37  0.60 42.48 Cl    
BMW135321.7+402151 13 53 21.67 +40 21 50.7 p		   PGC 0049347 NGC 5350                      SBbc   57  11.87    1.38  0.40   0.51  0.15 40.26 Gr 
BMW140715.2-270922 14 07 15.27 -27 09 27.7 e  30           PGC 0050373 ESO 510-G66                   E-S0  146  13.97    9.04  0.77   4.43  0.38 42.03 Cl
BMW141319.7-030857 14 13 19.76 -03 08 58.0 p               PGC 0050786 NGC 5507                      S0     43  13.22    0.31  0.09   0.14  0.04 39.47
BMW144104.0+532011 14 41 04.03 +53 20 10.7 p               PGC 0084264                                     629  17.53    5.88  1.02   2.18  0.38 42.99       
BMW151643.2+552432 15 16 42.26 +55 24 22.2 p               PGC 0054522 NGC 5908                      Sb     66  12.20    0.53  0.20   0.20  0.07 39.92
BMW160434.7+174315 16 04 34.65 +17 43 21.9 e 300           PGC 0056962 NGC 6041                      E-S0  209  13.90  183.30  4.47  80.65  1.97 43.60 Cl    
BMW160435.9+174325                        		   	        		      	     	              	        
BMW160508.1+174528 16 05 07.32 +17 45 27.4 p               PGC 0057031 NGC 6045                      SBc   204  13.58    1.12  0.25   0.49  0.11 41.37 Cl 
BMW160508.2+174340 16 05 08.28 +17 43 44.9 p               PGC 0057033 NGC 6047                      E     188  14.10    6.43  0.52   2.83  0.23 42.05 Cl
BMW160508.5+174344                        		   	        		      	     	              	        
BMW160533.6+173611 16 05 32.96 +17 35 58.8 e  40           PGC 0057062 IC 1178                       E-S0  204  14.74    2.57  0.53   1.13  0.24 41.71 Cl
BMW160534.4+323942 16 05 34.64 +32 39 37.4 p               PGC 0057078 CGCG 195-013                  Sc         15.29    2.12  0.44   0.87  0.18       c
BMW160544.5+174303 16 05 44.49 +17 43 02.4 p               PGC 0057096 IC 1185                       Sab   208  14.47    5.61  0.51   2.47  0.23 42.09 Cl
BMW161533.6-603948 16 15 32.22 -60 39 55.3 p               PGC 0057637 ESO 137-G7                    E-SO  100  13.08    8.95  1.48   8.33  1.38 41.96 Cl    
BMW161546.1-605509 16 15 46.04 -60 55 08.7 p               PGC 0057649 ESO 137-G8                    E      77  12.28    6.43  0.71   5.98  0.66 41.60 Cl
BMW161720.2+345408 16 17 20.23 +34 54 07.9 e  20           PGC 0057728 NGC 6107                      E     184  14.26    4.32  0.45   1.77  0.18 41.84 c; Gr
BMW165944.0+323656 16 59 44.21 +32 36 58.2 e  50           PGC 0059392                                     591  15.71   30.20  3.62  12.38  1.48 43.71 Cl    
BMW171305.9+434202 17 13 05.49 +43 42 02.2 p               PGC 0059859 CGCG 225-70                         330  15.75    4.68  0.55   1.92  0.22 42.37
BMW171519.1+573932 17 15 18.99 +57 39 33.5 p               PGC 0059942 MCG 10-24-118                       168  16.13    4.99  0.72   2.20  0.32 41.85
BMW173356.4+433936 17 33 56.58 +43 39 30.9 p               PGC 0060505 CGCG 226-29                              16.31    1.84  0.49   0.81  0.21          
BMW174336.7+330457 17 43 37.14 +33 04 57.9 p               PGC 0060757 CGCG 199-10                              15.73    4.30  0.85   2.02  0.40         
BMW181055.5+500049 18 10 54.76 +50 00 48.0 p               PGC 0061505 MCG 08-33-27                  E-S0  296  14.85    1.24  0.41   0.58  0.19 41.77 Cl?
BMW184359.5+453851 18 43 59.70 +45 38 55.0 e  30           PGC 0165780                                     263  16.00    1.22  0.42   0.62  0.21 41.67
BMW190012.7-362702 19 00 12.47 -36 27 04.6 p               PGC 0209730 CGMW 4-4359                                       4.03  0.98   2.22  0.54             
BMW192131.5+435941 19 21 31.81 +43 59 41.4 p               PGC 0063122 CGCG 230-9                    Sc    273  13.81   10.83  1.75   5.96  0.96 42.70 c; Cl 
BMW201127.3-564408 20 11 27.57 -56 44 05.5 p               PGC 0064228                               E     324  15.74    8.58  0.59   4.20  0.29 42.70 Cl
BMW201214.3-563822 20 12 14.71 -56 38 16.4 p               PGC 0095814                                          17.37    5.05  0.50   2.48  0.25       Cl? 
BMW204843.3+251650 20 48 43.29 +25 16 49.8 p               PGC 0065435 MCG 04-49-3                   E-S0  291  17.00    3.62  0.46   2.10  0.27 42.30 Cl
BMW210716.0-252807 21 07 16.01 -25 28  7.2 p               PGC 0066136 NGC 7016                      E     219  14.28   10.18  1.04   4.99  0.51 42.42 Cl 
BMW210720.4-252917 21 07 20.05 -25 29 16.2 p               PGC 0066137 NGC 7017                      SO    222  14.24    4.18  0.83   2.05  0.41 42.07 Cl 
BMW215209.2-194323 21 52 09.58 -19 43 22.7 e  20           PGC 0067523 ESO 600-G14                   S0-a  581  15.36    2.12  0.45   0.93  0.20 42.54 C?
BMW215209.4-194318                        		   	        		      	     	              	        
BMW221818.1-593625 22 18 18.55 -59 36 23.4 p               PGC 0068536 IC 5187                       Sc         14.86    5.57  1.08   2.28  0.44       c     
BMW223633.3+343010 22 36 32.97 +34 30 09.2 p               PGC 0069291                                     174 (16.60)   1.30  0.36   0.72  0.20 41.38
BMW223718.9+342654 22 37 19.17 +34 26 55.7 p               PGC 0069338 NGC 7335                      S0-a  126  13.90    1.10  0.32   0.60  0.18 41.03 c; Gr
BMW231221.4-435145 23 12 21.96 -43 51 53.2 p               PGC 0070684 ESO 291-G6                    E-S0       15.14    2.27  0.73   0.93  0.30       Cl?
BMW231546.1-590313 23 15 45.97 -59 03 11.3 p               PGC 0070861 ESO 148-IG2                   Sm    268  14.76    1.80  0.49   0.79  0.21 41.81 c
BMW231713.2+184229 23 17 11.37 +18 42 32.8 e  80           PGC 0070934 NGC 7578B                     E     241  14.21   43.44  2.43  21.28  1.19 43.14 Gr    
BMW231756.7-421339 23 17 56.23 -42 13 32.5 p               PGC 0131791                               S0-a  336  15.63    0.53  0.27   0.22  0.11 41.41 Gr?
BMW231855.6-421406 23 18 55.60 -42 14 05.9 e  30           PGC 0071031 NGC 7590                      Sbc    26  11.37    3.44  0.58   1.41  0.24 40.03 Gr 
BMW231855.6-421359                        		   	        		      	     	              	        
BMW231921.5-415910 23 19 21.75 -41 59 02.9 p               PGC 0095173                               Sa         16.00    2.12  0.65   0.87  0.27       Gr?
BMW231921.5-415906                        		   	        		      	     	              	        
BMW232042.7+081305 23 20 42.72 +08 13 05.1 p               PGC 0071140 NGC 7626                      E      69  11.92    5.55  1.05   2.72  0.51 41.16 Cl    
BMW233842.8+271300 23 38 42.74 +27 12 59.7 p               PGC 0071994 CGCG 476-95                   E     186  14.76    0.52  0.25   0.25  0.12 40.89 Cl
BMW234818.3+010623 23 48 18.80 +01 06 24.0 p               PGC 0196693                                     584  16.46    0.34  0.20   0.16  0.09 41.89
BMW234823.7-280427 23 48 23.98 -28 04 24.3 p               PGC 0072471 DUKST 471-12                  SO    203  14.64    3.61  0.87   1.48  0.36 41.84 Cl
BMW234847.1-281217 23 48 46.73 -28 12 18.8 p               PGC 0085790 DRCG54-31                     SO         16.14    5.17  1.22   2.12  0.50       Cl?   
BMW235415.0-100955 23 54 14.53 -10 09 52.0 e  55           PGC 0141285                               Sbc   394  16.29    3.75  0.62   1.65  0.27 42.47 Cl
BMW235415.1-100950                        		   	        		      	     	              	        
\end{verbatim}
\end{landscape}

\end{document}

%% file: tablecompl.tex
\begin{table*}
\begin{center}
\begin{tabular}{lcrclrcl}
\hline \hline
LEDA name & N$_H$ &\multicolumn{3}{c}{Count rate} & \multicolumn{3}{c}{F$_{0.5 - 2}$} \\
	  & (10$^{20}$ cm$^{-2})$ &\multicolumn{3}{c}{(10$^{-3}$ count s$^{-1})$} & \multicolumn{3}{c}{(10$^{-13}$ \ergcmsec)} \\
\hline \hline
0157406 & 4.9 & 0.37 &$\pm$& 0.07 &  0.12 &$\pm$& 0.02 \\
0068536 & 2.5 & 1.11 &$\pm$& 0.34 &  0.32 &$\pm$& 0.09 \\
0025825 & 4.2 & 1.12 &$\pm$& 0.22 &  0.35 &$\pm$& 0.06 \\
0042833 & 2.0 & 1.31 &$\pm$& 0.23 &  0.35 &$\pm$& 0.05 \\
0028990 & 0.8 & 1.50 &$\pm$& 0.30 &  0.36 &$\pm$& 0.07 \\
0005323 & 3.8 & 1.33 &$\pm$& 0.28 &  0.41 &$\pm$& 0.09 \\
0010446 & 3.1 & 1.48 &$\pm$& 0.24 &  0.43 &$\pm$& 0.06 \\
0057078 & 2.3 & 1.63 &$\pm$& 0.28 &  0.44 &$\pm$& 0.08 \\
0069338 & 8.8 & 1.24 &$\pm$& 0.20 &  0.46 &$\pm$& 0.07 \\
0038773 & 2.0 & 1.70 &$\pm$& 0.11 &  0.46 &$\pm$& 0.03 \\
0001333 & 4.2 & 1.55 &$\pm$& 0.27 &  0.48 &$\pm$& 0.09 \\
0007289 & 5.5 & 1.71 &$\pm$& 0.26 &  0.56 &$\pm$& 0.10 \\
0070861 & 2.6 & 2.11 &$\pm$& 0.38 &  0.61 &$\pm$& 0.12 \\
0018991 & 5.4 & 2.18 &$\pm$& 0.32 &  0.70 &$\pm$& 0.10 \\
0004117 & 4.0 & 2.36 &$\pm$& 0.23 &  0.73 &$\pm$& 0.06 \\
0043675 & 3.9 & 2.49 &$\pm$& 0.37 &  0.77 &$\pm$& 0.12 \\
0002204 & 1.9 & 3.23 &$\pm$& 0.23 &  0.87 &$\pm$& 0.05 \\
0045318 & 1.0 & 3.78 &$\pm$& 0.60 &  0.91 &$\pm$& 0.14 \\
0046432 & 3.2 & 3.13 &$\pm$& 0.66 &  0.91 &$\pm$& 0.20 \\
0047432 & 2.3 & 3.61 &$\pm$& 0.64 &  0.97 &$\pm$& 0.16 \\
0130966 & 2.4 & 4.41 &$\pm$& 0.81 &  1.19 &$\pm$& 0.22 \\
0016574 & 5.9 & 3.73 &$\pm$& 0.51 &  1.23 &$\pm$& 0.17 \\
0005324 & 3.8 & 4.34 &$\pm$& 0.51 &  1.35 &$\pm$& 0.16 \\
0057728 & 1.5 & 5.11 &$\pm$& 0.39 &  1.38 &$\pm$& 0.11 \\
0013368 & 1.4 & 7.22 &$\pm$& 1.41 &  1.73 &$\pm$& 0.24 \\
0029050 & 0.8 & 10.20 &$\pm$& 0.70 &  2.45 &$\pm$& 0.17 \\
0002362 & 2.7 & 9.06 &$\pm$& 0.99 &  2.63  &$\pm$& 0.29 \\
0004848 & 6.1 & 9.26 &$\pm$& 2.43 &  3.06 &$\pm$& 0.66 \\
0063122 & 7.9 & 11.02 &$\pm$& 1.42 &  3.97 &$\pm$& 0.36 \\
0006367 & 3.0 & 14.80 &$\pm$& 0.60 &  4.29 &$\pm$& 0.17 \\
0017451 & 7.5 & 15.66 &$\pm$& 1.72 &  5.64  &$\pm$& 0.72 \\
0028995 & 4.8 & 34.20 &$\pm$& 2.00 &  10.94 &$\pm$& 0.64 \\
\hline \hline
\end{tabular}
\end{center}
\caption{Galactic column density (from Dickey \& Lockman 1990), BMW count rates and fluxes for
galaxies in the complete serendipitous sample, ordered by increasing flux.}
\label{compltab}
\end{table*}